\documentclass[prx,aps,floats,twocolumn,longbibliography]{revtex4-1}

\usepackage[dvips]{graphicx}
\usepackage{amssymb}
\usepackage{amsmath}
\usepackage{natbib}
\usepackage{color}
\usepackage{chngcntr}
\usepackage{enumitem}
\usepackage{hyperref}

\hypersetup{
colorlinks=false,
linktocpage=true,
citebordercolor=[rgb]{0.5,0.5,1.0},
linkbordercolor=[rgb]{1.0,0.5,0.5},
urlbordercolor=[rgb]{0.5,0.5,1.0}
}

\begin{document}

\title{Classical Dimers on Penrose Tilings}

\author{Felix Flicker} 
\email{flicker@physics.org}
  \affiliation{Rudolf Peierls Center for Theoretical Physics, Oxford OX1 3PU, United Kingdom}
\author{Steven H. Simon}
\email{steven.simon@physics.ox.ac.uk}
  \affiliation{Rudolf Peierls Center for Theoretical Physics, Oxford OX1 3PU, United Kingdom}
\author{S. A. Parameswaran}
\email{sid.parameswaran@physics.ox.ac.uk}
  \affiliation{Rudolf Peierls Center for Theoretical Physics, Oxford OX1 3PU, United Kingdom}
  
\begin{abstract}
We study the classical dimer model on rhombic Penrose tilings, whose edges and vertices may be identified with those of a bipartite graph. We find that Penrose tilings do not admit perfect matchings (defect-free dimer coverings). Instead, their {\it maximum matchings} have a monomer density of $81-50\varphi\approx 0.098$ in the thermodynamic limit, with $\varphi=\left(1+\sqrt{5}\right)/2$ the golden ratio. Maximum matchings divide the tiling into a fractal of nested closed regions bounded by loops that cannot be crossed by monomers. These loops connect second-nearest neighbor even-valence vertices, each of which lies on such a loop. Assigning a charge to each monomer with a sign fixed by its bipartite sub-lattice, we find that each bounded region has an excess of one charge, and a corresponding set of monomers, with adjacent regions having opposite net charge. The infinite tiling is charge neutral. We devise a simple algorithm for generating maximum matchings, and demonstrate that maximum matchings form a connected manifold under local monomer-dimer rearrangements. We show that dart-kite Penrose tilings feature an imbalance of charge between bipartite sub-lattices, leading to a minimum monomer density of $\left(7-4\varphi\right)/5\approx 0.106$ all of one charge.
\end{abstract}

\maketitle 

%
\section{Introduction}
%

Dimer models provide abstract yet solvable frameworks enabling mathematically precise statements applicable to a wide range of physical situations. The \emph{quantum} dimer model was introduced as an approximate treatment of fluctuating nearest-neighbor spin singlets in the resonating-valence-bond (RVB) state~\cite{RokhsarKivelson88,KivelsonEA87,MoessnerSondhi01,MoessnerRaman07}, proposed as a possible explanation for high-temperature superconductivity~\cite{FazekasAnderson74,Anderson87}. Describing spin configurations in terms of singlets naturally implies a hard-core constraint: a single site with a spin-$1/2$ degree of freedom can belong to at most one singlet, defining a dimer model. Although the spin-dimer mapping is not one-to-one, the dimer model is nevertheless a useful caricature of the underlying spins, and an intuitive understanding in terms of dimers often translates fruitfully to spin models despite their more complicated structure.

Work on dimer models also underpins research into topologically ordered states of matter~\cite{MoessnerSondhi01,MoessnerRaman07}. A defining characteristic of such phases is fractionalization, a phenomenon whereby the emergent excitations of a system appear as fractions of the microscopic degrees of freedom~\cite{SavaryBalents17}. In addition to its fundamental significance, fractionalization is also relevant to applications in which fractionalized quasiparticles can be used to perform quantum computation in a topologically protected manner~\cite{NayakEA08}. Dimer models provide a particularly elegant framework within which to study such phenomena~\cite{MoessnerSondhiFradkin}. Monomers --- obtained by breaking apart dimers, and hence fractionalized in an intuitive sense --- can be thought of as sources and sinks of an emergent gauge field. Quantum fluctuations (resonances) between dimer configurations give the gauge field dynamics. Depending on the lattice structure and dimensionality, at long wavelengths the gauge field dynamics can either exhibit confinement, or else can be described by a discrete or continuous gauge structure and, correspondingly, host gapped or gapless excitations~\cite{MoessnerSondhi01,MoessnerSondhiRVB_3D,MoessnerRaman07}. In both cases monomers may be separated to arbitrary distances at finite energy cost (they are \emph{deconfined}): the system thus hosts emergent fractionalized quasiparticles.

\begin{figure}[t!]
\includegraphics[width=.4\textwidth]{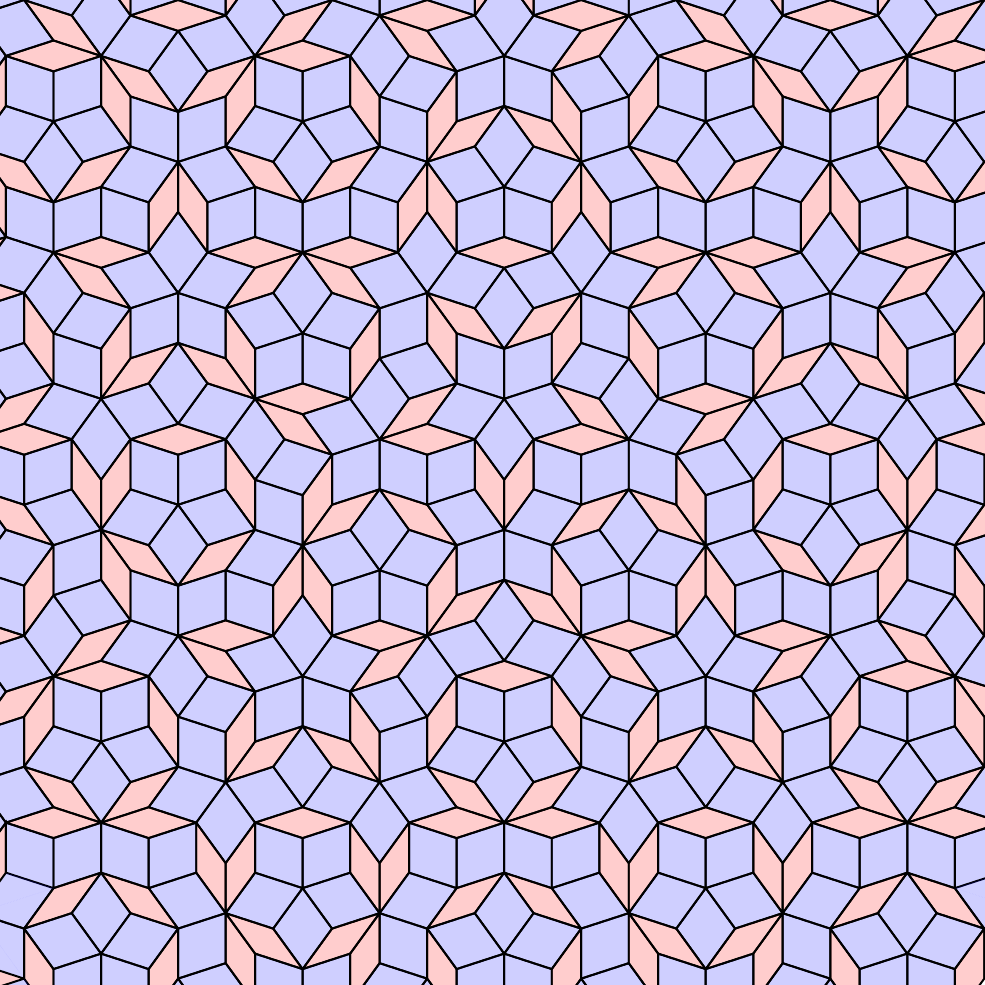}
\caption{A finite section of the Penrose tiling constructed of two rhombuses (colored red and blue here).
}
\label{fig:tiling}
\end{figure}

The understanding of quantum dimer models frequently draws on highly influential exact results on their classical counterparts~\cite{Kasteleyn61,Fisher61,FisherTemperley61,Kasteleyn67,HurstGreen60}. Insights are also afforded by numerical simulations of classical dimers~\cite{ClassicalDimer3D,KrauthMoessner03}, that are often more computationally tractable than their quantum generalizations. Studying dimer coverings of graphs remains an active area of current research in mathematics and statistical physics~\cite{RichardEA98,KenyonEA00,Kenyon02,KenyonOkounkov06,KenyonOkounkov07,DijkgraafEA07}. For both the classical and quantum cases, results to date have focussed primarily on periodic graphs, partly because of their relative simplicity and the resulting potential for exact results, and partly because of the relevance to physical systems such as crystal lattices~\cite{Kasteleyn61,KenyonEA00}. There is also an active interest in investigating dimers on random graphs, such as those with quenched disorder~\cite{KrapivskyEA,ErdosRenyi,NewmanEA01,AlbericiEA15}. 

Traditionally these two extremes, periodicity and disorder, were the only cases studied in materials physics. This changed with the discovery of \emph{quasicrystals}, states of matter with properties intermediate between the periodic order of crystals and the disorder of glasses. The identification of quasicrystals via their diffraction patterns --- which feature discrete rotational symmetries forbidden in periodic crystals --- led to a redefinition of crystallography in the second half of the twentieth century~\cite{Shechtman84}. Perhaps the simplest route to understanding quasicrystals is through aperiodic tilings such as the Penrose tiling (Fig.~\ref{fig:tiling}), which lack the discrete translational invariance of periodic lattices featuring instead a discrete scale invariance~\cite{Penrose74,Senechal,Janot,GrunbaumShephard}. Quasicrystals are real materials with the symmetries of Penrose-like tilings, just as crystals are real materials with the space-group symmetries of periodic lattices~\footnote{We use the phrase \emph{Penrose-like tiling} to indicate aperiodic tilings featuring diffraction patterns (Fourier transforms) with Bragg peaks and dense backgrounds which feature rotational symmetries forbidden by crystallographic restriction. We use the phrase \emph{quasilattice} for their one-dimensional equivalents.}. Although a large body of work has explored single-particle phenomena in quasiperiodic systems~\cite{SocolarEA86,KohmotoEA87,AshraffEA90}, including more recent extensions to incorporate topological properties~\cite{TopoQC,MadsenEA13,FlickerVanWezel15}, few studies have explored strongly correlated phenomena in quasicrystals. Recent interest in understanding many-body localization in quasiperiodic systems~\cite{IyerEA13,schreiber2015observation,MBLQPTransition}, as well as the relevance of quasicrystals to magnetic insulators, heavy fermion materials~\cite{Jagannathan12,ThiemChalker15A,ThiemChalker15B,Andrade15,Sato_2017}, and even superconductivity~\cite{Kamiya:2018aa}, suggest that the time is ripe to investigate such problems.

Here, we combine these two distinct lines of investigation and extend the study of dimer models to include quasiperiodic graphs. The reason for studying dimer models is twofold: on the one hand, they account for the physics of magnetic frustration and local constraints (textbook correlation effects) from the outset, and, on the other, our analysis can leverage insights from combinatorial graph theory. Specifically,  we consider classical dimers on Penrose tilings, with the vertices and edges of the tiling considered the vertices and edges of a graph. Perhaps unsurprisingly, this case proves fundamentally distinct from both periodic and disordered graphs. We prove a number of exact results. For the majority of the paper we consider the Penrose tiling constructed from two rhombic tiles shown in Fig.~\ref{fig:tiling}. For this system we prove that it is not possible to achieve a perfect matching of dimers to vertices, such that each vertex touches precisely one dimer. We then turn to maximum matchings, in which the maximum number of dimers appears in the graph, with no vertex connecting to two dimers. We prove that the density of monomer defects, vertices not reached by dimers, is $81-50\varphi$, with $\varphi=\left(1+\sqrt{5}\right)/2$ the golden ratio. We provide an algorithm for generating maximum matchings. Considering the monomers as mobile particles, with motion defined by a local reconfiguration of dimers, we prove that monomers are restricted to closed, finite regions of the graph, which appear in a nested fractal structure. We prove that maximum matchings form a manifold connected by local monomer-dimer moves. Turning briefly to the wider class of Penrose-like tilings, we prove that a variation on the Penrose tiling, made instead from tiles shaped as \emph{darts} and \emph{kites}, is also unable to admit perfect matchings. We prove that the minimum monomer density is precisely $\left(7-4\varphi\right)/5$ in this case, and that all monomers are of the same bipartite charge. Considering aperiodic tilings other than Penrose we provide evidence in support of our conjecture that certain examples admit perfect matchings. On the other hand, we prove that certain other examples cannot admit perfect matchings. We demonstrate that these latter cases feature broadly similar behavior to the rhombic Penrose tiling.

\begin{figure}[t]
\includegraphics[width=.4\textwidth]{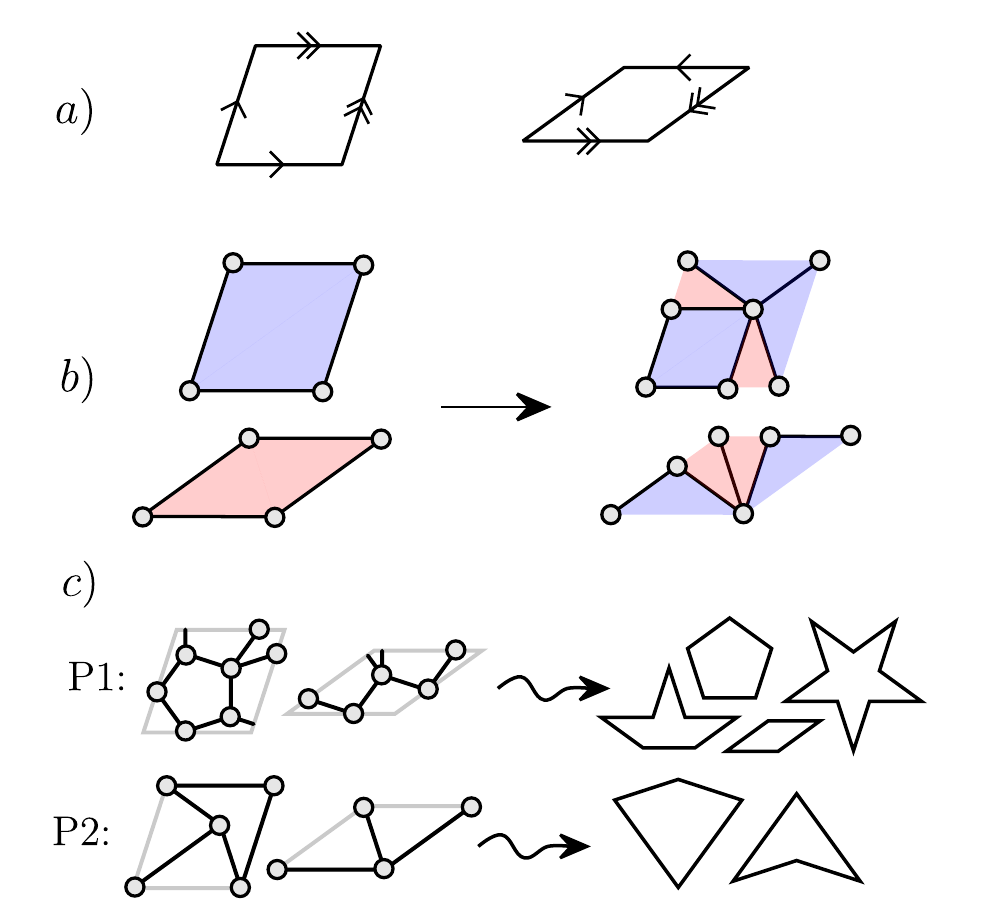}
\caption{$(a)$ The Penrose tiling can be created by decorating the rhombuses with \emph{matching rules}, where the decorations of neighboring edges must match. $(b)$ An alternative method of creating the tiling uses \emph{inflation rules}, in which each tile is subdivided into a combination of the two tiles as shown. Black lines indicate graph edges, and grey circles graph vertices. $(c)$ Decorating the tiles leads to different Penrose tilings made from different tile types: the P1 pentagonal tiling, and the P2 dart-kite tiling.
}
\label{fig:tiles}
\end{figure}

This paper proceeds as follows. In Section~\ref{sec:background} we provide background on Penrose tilings and dimer matchings of graphs. In Section~\ref{sec:no_perfect_matchings} we prove that Penrose tilings do not admit perfect matchings, \emph{i.e.}~they must feature a finite density of monomer defects, and study properties of the boundaries which restrict the movement of monomers. In Section~\ref{sec:maximum_matchings} we provide an algorithm for generating maximum matchings. In Section~\ref{sec:monomer_densities} we find the minimum density of monomers in the infinite Penrose tiling analytically, and numerically confirm the result. In Section~\ref{sec:Ergodicity} we demonstrate that maximum matchings form a manifold connected by local monomer-dimer moves. In Section~\ref{sec:other_quasilattices} we consider classical dimers on other Penrose-like tilings. Finally, in Section~\ref{sec:conclusions} we provide concluding remarks.

%
\section{Background}
\label{sec:background}
%

%
\subsection{Penrose Tilings}
\label{subsec:background_penrose}

Penrose tilings are aperiodic covers of the Euclidean plane by sets of inequivalent tiles~\cite{Penrose74,Senechal,Janot}. Throughout most of this paper we take as the set two rhombuses (the so-called P3 tiling, shown in Fig.~\ref{fig:tiling}). Other Penrose tilings can be created as decorations of the P3 tiling, as shown in Fig.~\ref{fig:tiles}c; one such example, the P2 tiling whose tiles are darts and kites, we consider in Section~\ref{sec:other_quasilattices}. Unless otherwise stated, `Penrose tiling' will be assumed to mean the rhombic tiling. Penrose tilings lack the discrete translational invariance of periodic lattices, featuring instead a discrete scale invariance.  They were originally devised as a problem in recreational mathematics, extending attempts to create aperiodic tilings begun by Kepler~\cite{Kepler}. The study of Penrose tilings became relevant to physics with the observation that certain alloys demonstrate closely-related symmetries~\cite{Shechtman84}. The resulting \emph{quasicrystals} can be identified by their diffraction patterns, which feature 5-, 8-, 10-, or 12-fold rotational symmetries~\cite{Levitov88}, in violation of the crystallographic restriction theorem which states that periodic structures in 2D or 3D can feature only 2-, 3-, 4-, or 6-fold rotations~\cite{AshcroftMermin}. 

\begin{figure}[t]
\includegraphics[width=.45\textwidth]{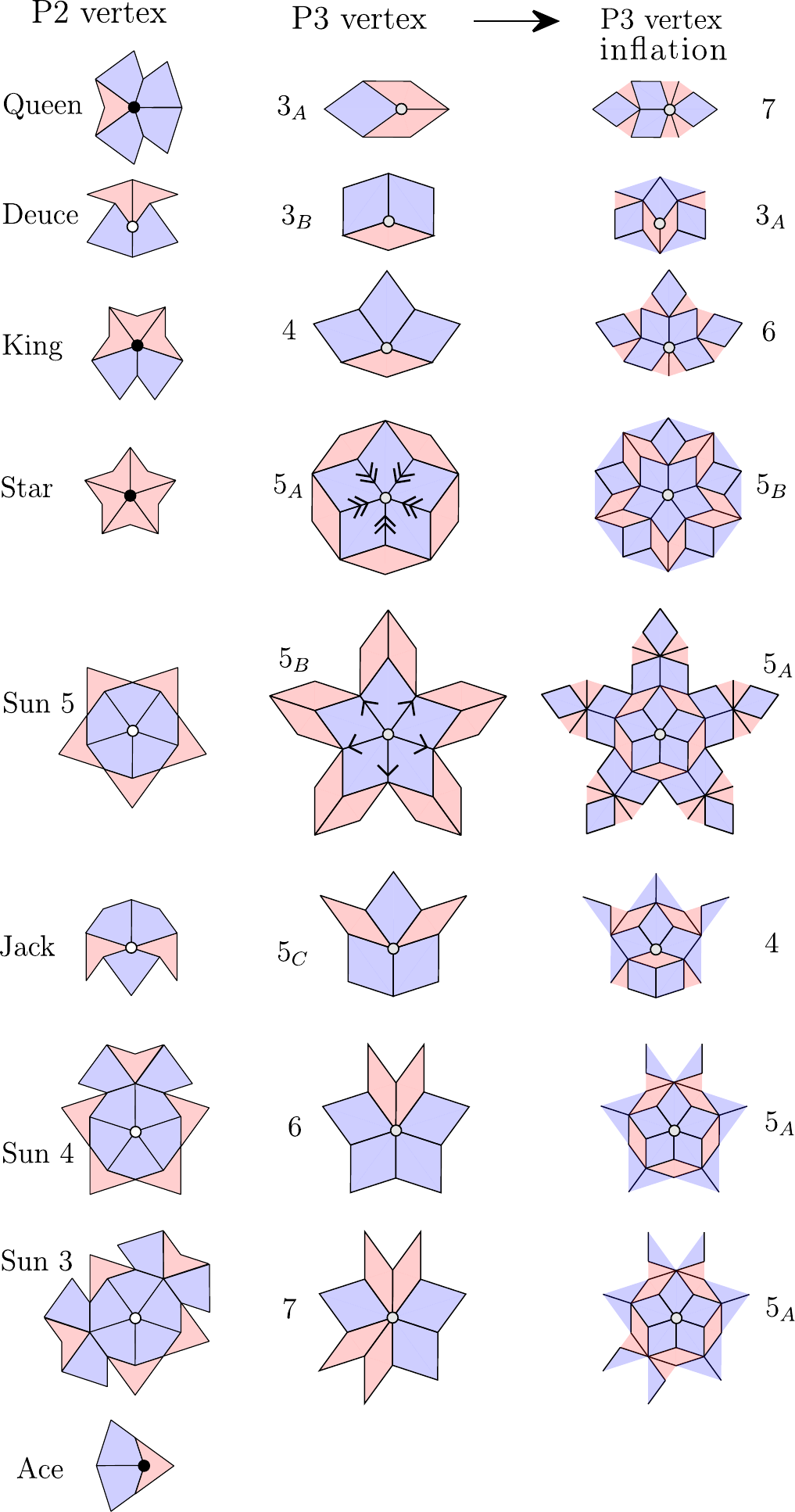}
\caption{The possible vertices in the P2 dart-kite tiling (left), their equivalents in the P3 rhombic tiling (middle), and the inflations of each (right). We label the P3 vertices according to the number of edges connecting to them. Tile matching rules distinguish the $5_{A,B}$-vertices and are indicated; P2 Suns are labeled according to the number of darts connecting to them~\cite{Henley86}. The P2 vertices divide into two bipartite sub-lattices: Star-Queen-King-Ace (black vertices) and Deuce-Jack-Suns (white vertices). 
}
\label{fig:vertices}
\end{figure}

The Penrose tiling can be composed of the two rhombuses shown in Figs.~\ref{fig:tiling} and \ref{fig:tiles}. The two rhombuses have internal angles (as multiples of $2\pi/10$) $\left\{2,3\right\}$ (\emph{thick}, blue in the figures) and $\left\{1,4\right\}$ (\emph{thin}, red in the figures). In order to force the tiling to be aperiodic, so-called matching rules must be applied to the tiles (Fig.~\ref{fig:tiles}a): decorations of the edges such that only like edges may meet in the tiling~\cite{deBruijn81A}. Any tiling obeying the matching rules is a Penrose tiling; however, starting from a finite seed and growing the tiling by locally adding new tiles at the edge, it is possible to reach arrangements in which the tiling cannot be grown any further~\cite{Senechal,Janot,GrunbaumShephard}. Fig.~\ref{fig:vertices} shows the eight possible ways in which the tiles can meet at a vertex, consistent with the matching rules.

An alternative, recursive, algorithm for generating Penrose tilings is illustrated in Fig.~\ref{fig:tiles}b. In this approach, rules are defined for sub-dividing each rhombus into a combination of the two rhombuses, in a process called \emph{inflation}~\cite{GrunbaumShephard}. The resulting combination is then rescaled so as to be constructed of exact copies of the original rhombuses (the rescaling is not shown here). The inflations of the eight vertex types are shown in Fig.~\ref{fig:vertices}. An infinite number of inflations applied to any finite simply-connected tile set constructed from the tiles in Fig.~\ref{fig:tiles} results in a Penrose tiling. This construction also makes apparent the discrete scale invariance of the Penrose tiling: inflation leads to a \emph{locally isomorphic} tiling, meaning that every finite set of tiles found in one can be found in the other~\cite{Gardner}. The presence of triangles (half-rhombuses) in the base inflation units leads to similar triangles on boundaries upon inflation, but these are negligible in the thermodynamic limit. Denoting the number of thick (blue) and thin (red) tiles after $n$ inflations as $b_n$ and $r_n$, respectively, their growth can be characterized by a $2\times2$ matrix: 
\begin{align}
\left(\begin{array}{c}
b_{n+1}\\
r_{n+1}
\end{array}\right)= \left(\begin{array}{cc}
2 & 1\\
1 & 1
\end{array}\right)\left(\begin{array}{c}
b_n\\
r_n
\end{array}\right).\end{align}
Under repeated applications of the matrix, \emph{i.e.}~repeated inflations of the tiling, we find
\begin{align}
\left(\begin{array}{c}
b_n\\
r_n
\end{array}\right)= \left(\begin{array}{cc}
2 & 1\\
1 & 1
\end{array}\right)^n\left(\begin{array}{c}
b\\
r
\end{array}\right)=\left(\begin{array}{c}
F_{2n+1}\,b+F_{2n}\,r\\
F_{2n}\,b+F_{2n-1}\,r
\end{array}\right)
\end{align}
where the initial numbers of tiles are $b$ and $r$. The growth of the number of tiles is controlled by the Fibonacci numbers $F_n$ (for $n\ge0$):
\begin{align}
n &=0,\,1,\,2,\,3,\,4,\,5,\,6,\,\,\,7,\,\,\,\,8,\,\,\,\,\,9,\,10,\,\ldots\nonumber\\
F_{n} &=0,\,1,\,1,\,2,\,3,\,5,\,8,\,13,\,21,\,34,\,55,\,\ldots 
\end{align}
The largest eigenvalue of the matrix is $\varphi^2$, with $\varphi=\left(1+\sqrt{5}\right)/2$ the golden ratio. $\varphi$ is a Pisot-Vijayaraghavan (PV) number: a number with modulus greater than one, whose Galois conjugates all have modulus strictly less than one~\cite{Senechal,BoyleSteinhardt16A,BoyleSteinhardt16B}. Any integer power of a PV number is also a PV number. All quasicrystals and Penrose-like tilings can be generated by inflation, and in all cases the largest eigenvalue of the inflation matrix is a quadratic-irrational PV number~\cite{BombieriTaylor86,ZaporskiFlicker18}. In the thermodynamic limit the largest eigenvalue dominates, and the ratio of the components of the associated right-eigenvector gives the ratio of the number of tiles of each type. For the Penrose tiling this ratio, of thick to thin tiles, is $\varphi$.

\begin{figure}[t]
\includegraphics[width=.45\textwidth]{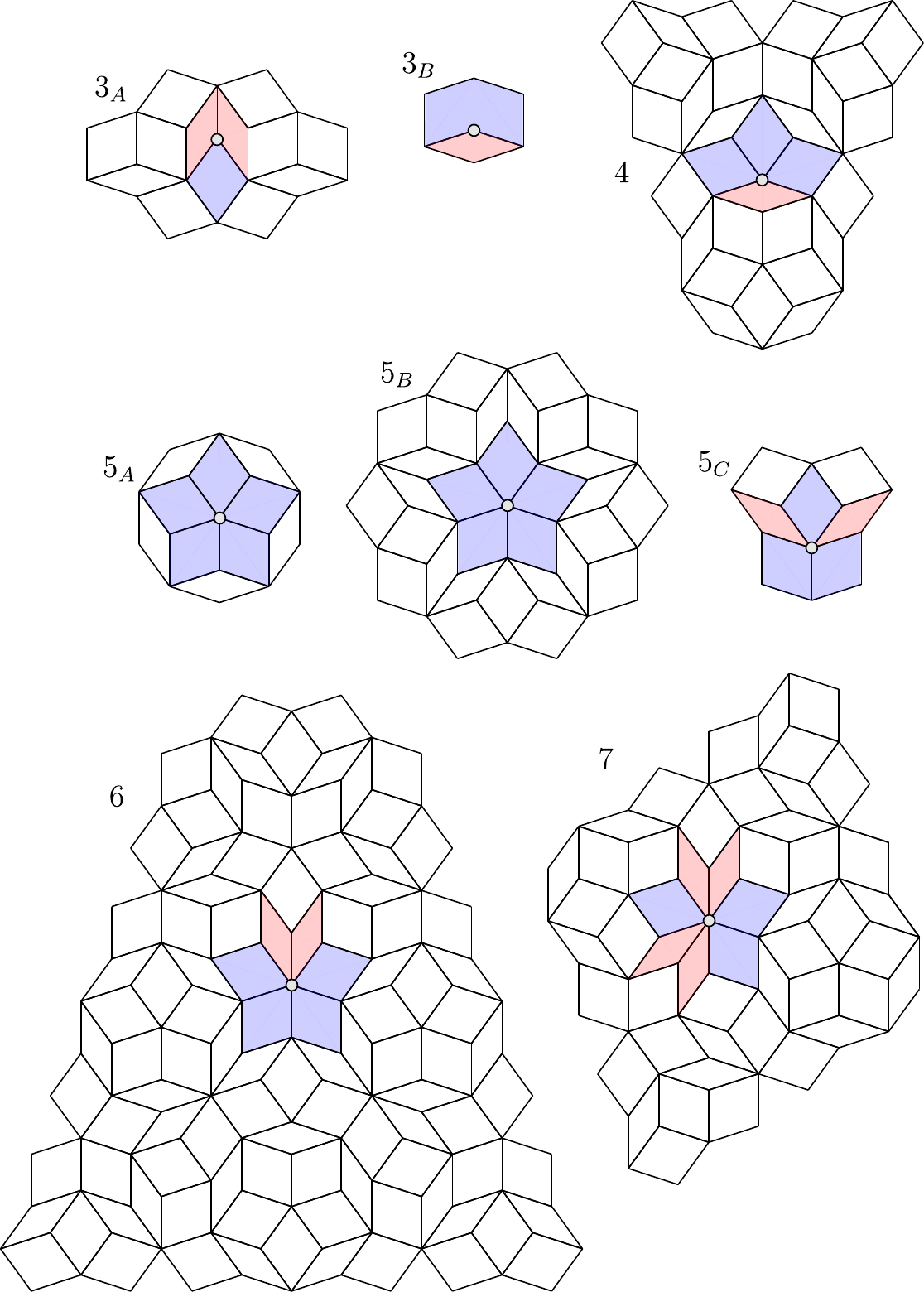}
\caption{Each vertex type in Fig.~\ref{fig:vertices} is accompanied by a set of tiles which always appears around it, termed a \emph{local empire}~\cite{GrunbaumShephard}.
}
\label{fig:empires}
\end{figure}

While the aperiodic nature of the tiling requires ambiguity in tile placements, the matching rules of Fig.~\ref{fig:tiles}a may force certain tile placements given certain local configurations. The set of tiles which unambiguously appears around a given feature is known as an \emph{empire}~\cite{GrunbaumShephard}. In general, the empire of a feature will not be simply connected; the set of tiles simply connected to the feature is known as the \emph{local empire}. Fig.~\ref{fig:empires} shows the local empires of each vertex type in the Penrose tiling~\cite{EffingerDean}. These will play a major role in our discussion, and crucially will enable us to prove general results about the entire tiling.

\subsection{Dimer Coverings}
\label{subsec:background_dimers}

There is an extensive literature on dimer coverings of graphs in both physics and mathematics. Here we present only the points salient to the remaining discussion in this paper; for a general introduction to the topic see Refs.~\onlinecite{Kasteleyn67,Gibbons,KenyonEA00}, and for an introduction to the relevance to physics see Refs.~\onlinecite{Baxter,Kenyon02}.

\begin{figure}[t]
\includegraphics[width=.4\textwidth]{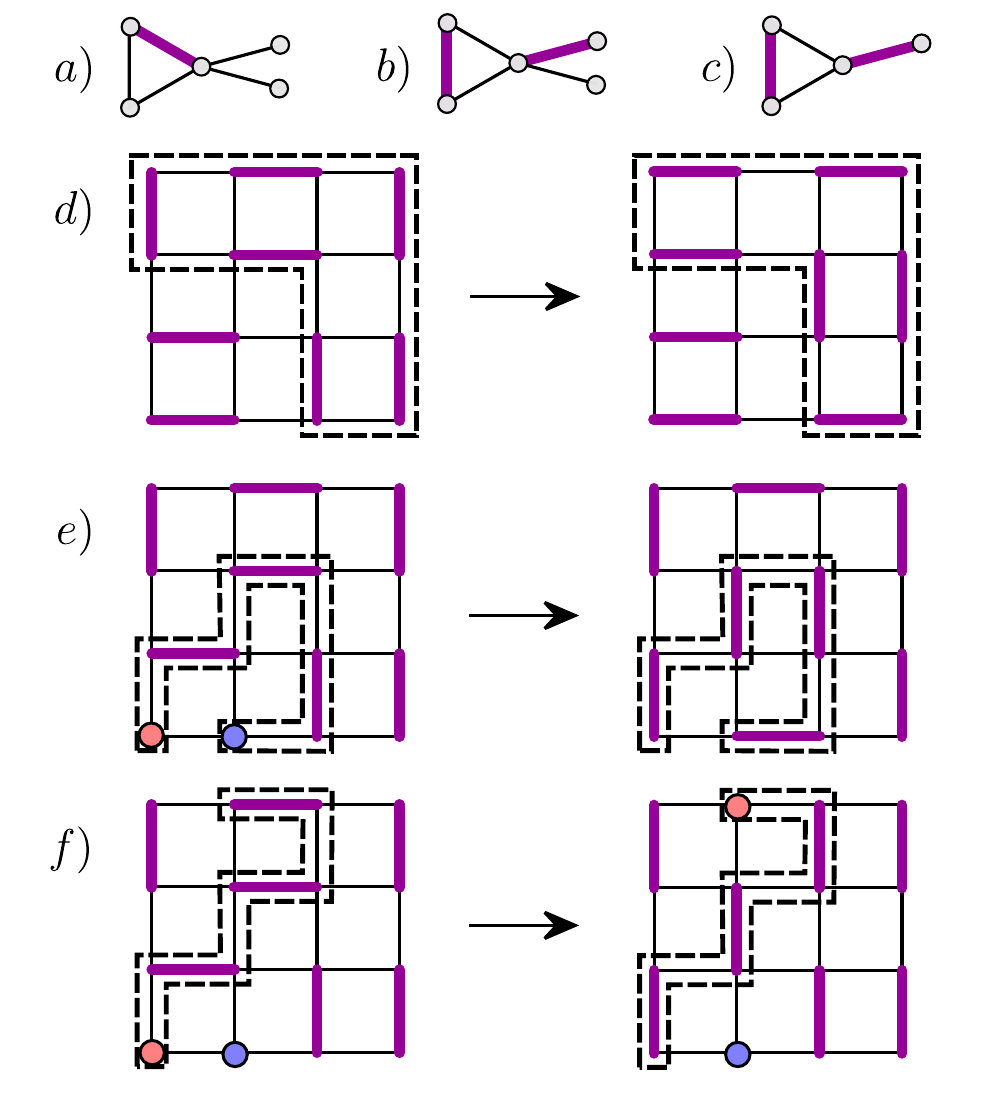}
\caption{$(a)$ A \emph{maximal} matching of a graph is a dimer covering such that no further edge can be covered with a dimer (purple) without causing a vertex (grey circles) to connect to two dimers. $(b)$ A \emph{maximum} matching additionally contains the maximum number of dimers. $(c)$ If every vertex in a maximum matching connects to a dimer, the result is a \emph{perfect matching}. $(d)$ An \emph{alternating cycle} is enclosed by the dashed line in this perfect matching of a bipartite graph. \emph{Augmenting} the cycle (switching which edges are covered by dimers) results in another perfect matching. $(e)$ Deleting one dimer has resulted in a monomer anti-monomer pair. \emph{Augmenting paths} are alternating paths with both ends terminating on monomers. Augmenting the path removes both monomers. No augmenting path can be present in a maximum matching. $(f)$ An alternating path terminating on one monomer is highlighted. Augmenting the path moves the monomer. We term a \emph{minimal monomer move} the augmentation of an alternating path of minimum length, which enacts a monomer hop across one unoccupied and one occupied edge.
}
\label{fig:graphs}
\end{figure}

A graph is a set of vertices connected by edges. We will consider planar graphs, which can be embedded in the Euclidean plane such that no edges pass under or over one another. The graphs we consider are also bipartite, meaning that the vertices can be colored one of two colors, say red and blue, such that edges only connect red vertices to blue, and vice versa (see Fig.~\ref{fig:graphs}). An equivalent statement is that all cycles (loops, or closed paths) on the graph are of even length~\cite{Gibbons}.

A \emph{matching} of a graph is a set of edges such that no vertex touches more than one edge in the set~\cite{Gibbons}. A matching is equivalent to a hard-core dimer covering, and henceforth we will use the terms interchangeably, and consider an edge to be covered by a dimer if it belongs to the matching. A \emph{maximal} matching is a matching in which no further edges can be covered by dimers while remaining a matching (Fig.~\ref{fig:graphs}a). We focus here on \emph{maximum} matchings, maximal matchings with the additional property that the dimers cover the maximum possible number of edges (Fig.~\ref{fig:graphs}b). A \emph{perfect} matching is a maximum matching in which {\it every} vertex connects to a dimer (Fig.~\ref{fig:graphs}c). There may be more than one perfect matching. Starting from a perfect matching of a bipartite graph (if one exists), deleting one dimer leads to two \emph{monomer} defects where vertices connect to zero dimers (Figs.~\ref{fig:graphs}d--f). A maximum dimer matching hosts the minimum number of monomers~\cite{Gibbons}. Various algorithms exist for generating maximum matchings of graphs, such as the Hopcroft-Karp algorithm for bipartite graphs~\cite{HopcroftKarp73}. Along an \emph{alternating path} edges are alternately covered and not covered by dimers~\footnote{Note that paths are distinct from walks: paths are walks that cannot self-intersect.}. If the path closes, it is an \emph{alternating cycle}. If it does not close, one end necessarily terminates on a monomer. {An \emph{augmenting} path is an alternating path with both ends terminating on monomers.} In a bipartite graph, two monomers connected by an augmenting path must be of opposite bipartite charge; we term one a monomer and the other an anti-monomer, by analogy to particles. To \emph{augment} a general alternating path is to switch which edges are covered by dimers, and which are not. These cases are shown in Figs.~\ref{fig:graphs}d--f.

The creation of two monomers can be seen as analogous to the creation of a particle-antiparticle pair, with the charge of the particle being its bipartite color (red or blue). Rearranging dimers can have the effect of moving monomers; specifically, augmenting an alternating path terminating on a monomer moves the monomer along the length of the path. We define a \emph{minimal monomer move} to be a hop across one unoccupied edge and one occupied edge, \emph{i.e.}~augmenting a shortest-length alternating path terminating on the monomer. As the monomer moves along the path it switches which edges are covered by the dimers. An augmenting path connecting a monomer to an anti-monomer can be thought of as a classical version of the Dirac strings connecting magnetic monopoles to their antimonopoles, required by gauge consistency~\cite{Dirac31}. In a quantum gauge theory these Dirac strings (which ensure the single-valuedness of the wavefunction of an electron passing around the string) would be gauge-dependent quantities, requiring a precise relationship between the electric and magnetic charge quanta in order to be unobservable. This ambiguity can survive in the graph setting in the following sense: presented with a configuration of monomers and anti-monomers, it may not be possible to make an unambiguous statement as to where the strings of flipped dimers lie. The dimers in a liquid phase can be seen as a structured vacuum in which monomers move.

In this paper we identify the edges and vertices of the Penrose tiling with the edges and vertices of a graph, and seek properties of dimer coverings of this graph. While we are unaware of previous work on this subject, some results on more general planar bipartite graphs can be applied to this case, as in Refs.~\onlinecite{Mercat01,Kenyon02,deTiliere07}. Interacting spins on Penrose-like tilings were considered in Refs.~\onlinecite{Korepin87,YangPerk07,ThiemChalker15A,ThiemChalker15B}. Dimers on one-dimensional quasilattices are discussed in Ref.~\onlinecite{DunlapEA90}. Treated as a graph in this way, the Penrose tiling of course admits a planar embedding, and we will maintain the geometry of the tiling in the remaining discussion (although the results presented depend solely on the network topology of the graph). As the faces are all rhombuses, it follows that the graph is also bipartite: this is true for all planar graphs whose faces all have an even number of edges~\cite{Kempe79,Soifer}.

%
\section{Impossibility of Perfect Matchings on Penrose Tilings}
\label{sec:no_perfect_matchings}
%

Considering that the number of edges emanating from a vertex can range from three to seven in the Penrose tiling, it may not seem surprising that the corresponding graph does not admit a perfect matching. On the other hand, the graph is bipartite, and all cycles are of even length. The tiling can be constructed as a 2D slice through a 5D simple-hypercubic lattice; the graph equivalent of this higher-dimensional lattice would admit perfect matchings~\cite{Senechal,Janot}.

Finite sections of Penrose tilings can be created by inflating simply-connected sets of tiles using the inflation rules in Fig.~\ref{fig:tiles}. Maximum dimer matchings can then be found, for example, using the Hopcroft-Karp algorithm. However, monomers resulting from this process may in principle be able to hop to the boundary, by a sequence of minimal monomer moves (\emph{i.e.}~by augmenting alternating paths connecting each monomer to the boundary). In this case it would be unclear whether the monomers were an artefact of considering only a finite section of the tiling.

One statement about the Penrose tiling, however, follows straightforwardly from its bipartite structure: namely, that \textbf{any matching on the rhombic Penrose tiling is charge neutral in the thermodynamic limit}. \emph{Proof}: to see this, first note that the trivial matching, with zero dimers (monomers on all vertices), is charge neutral. This is because the average number of edges protruding from a vertex (the vertex \emph{valence}) is the same for both bipartite sub-lattices. The average valence across the entire graph must be four, as all tiles have four edges. The average valence of the two bipartite sub-lattices must be equal because all vertex types appear in both sub-lattices (easily checked in a finite region). Then note that any matching is formed by placing dimers onto this trivial matching. Each dimer eliminates one monomer of each bipartite charge, and therefore conserves total charge. $\square$

We now prove that a perfect matching of the Penrose tiling is not possible, as the structure of the tiling leads to closed loops of edges which can never be covered by dimers in maximum matchings. These loops cannot be crossed by alternating paths in maximum matchings, and so act as impenetrable obstacles to monomer movement, when the movement is defined via minimal monomer moves. The existence of a net imbalance of bipartite charge enclosed by such a loop then suffices to demonstrate the impossibility of perfect matchings.

\subsection{Impermeable Monomer Membranes}
\label{subsec:even_loop_existence}

\begin{figure}[t]
\includegraphics[width=.4\textwidth]{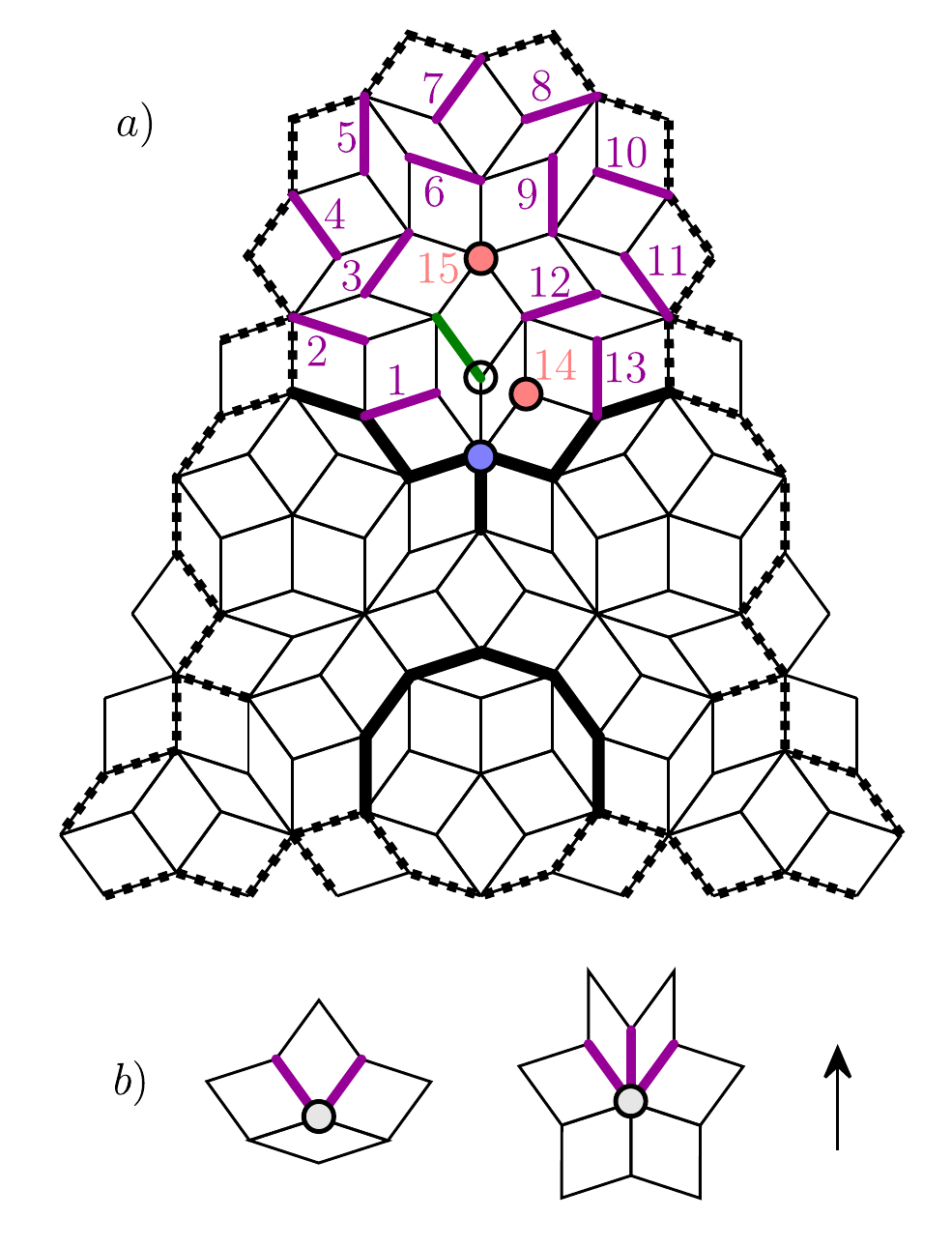}
\caption{Elements used in the proof that $(a)$ a $6$-vertex (blue) cannot host a monomer in a maximum matching, and $(b)$ the dimer which must emanate from the $6$-vertex must be placed on one of the three purple edges indicated. The direction of 4- and 6-vertices is defined to align with the arrow. Thick black lines indicate edges which cannot be covered by dimers in maximum matchings. Solid (dashed) lines indicate definite (potential) uncoverable edges. Potential uncoverable edges are those that may or may not become uncoverable depending on how the local empire overlaps with others in the full tiling.
}
\label{fig:even_loop_proof}
\end{figure}

As a starting observation, note that \textbf{any even-valence vertex in the Penrose tiling (a 4-vertex or 6-vertex) has no even-valence neighbors, and precisely two even-valence second-nearest neighbors} (see Fig.~\ref{fig:tiling})~\footnote{Note that we use the term nearest/next-nearest neighbor, \emph{etc.} in the sense of graph connectivity rather than actual spatial distance.}. \emph{Proof}: this can be proven by considering the local empire of the 6-vertex in Fig.~\ref{fig:empires}. This region is large enough to cover the entire tiling, allowing for overlaps~\cite{Gummelt96}. In Fig.~\ref{fig:even_loop_proof} the local empire of the 6-vertex is shown; the solid thick black lines connect second-nearest neighbor even-valence vertices (note that we include an extra twig on the 6-vertex; this will be explained shortly). The thick dashed black lines represent potential completions of the loop, which could become solid depending on how the incomplete boundary vertices are finished (\emph{i.e.}~the precise way in which each vertex appears in the tiling). The 6-vertex has zero even-valence neighbors, and two 4-vertices are the only even-valence second-nearest neighbors. Each of these 4-vertices has no even-valence neighbors, and, aside from the 6-vertex, may have either a 4-vertex or 6-vertex as another second-nearest neighbor (by considering all possible completions of the boundary vertices). Elsewhere in the local empire an arc of three second-nearest neighbor 4-vertices appears. Again considering all possible completions of boundary vertices, the only possible continuations are to a $-4-6-4-4-4-6-4-$ configuration, where only second-nearest neighbor even-valence vertices are listed, or a $\left(4-\right)^5$ configuration (we term this configuration a `$4^5$ loop'; we later show that it is the only exception to various rules). This completes the proof that even-valence vertices have zero even-valence vertices as neighbors and precisely two even-valence vertices as second-nearest neighbors. $\square$

A corollary of this result is that these chains of even-valence vertices must either form closed loops or cross the entire system. We explain this fact and consider further properties of the loops in Section~\ref{subsec:monomer_membrane_properties}. Here we demonstrate that they act as impermeable barriers to monomers within the set of maximum matchings. First, note that \textbf{even-valence vertices never host monomers in maximum matchings}. \emph{Proof}: this can be proven by a graphical argument shown in Fig.~\ref{fig:even_loop_proof}, which we now describe. The simplest case to prove is again that of the 6-vertex, owing to the large size of its local empire.  We begin by assuming that a monomer does exist on the 6-vertex (the blue vertex in Fig.~\ref{fig:even_loop_proof}a) and prove a contradiction: the monomer always connects via an augmenting path to an anti-monomer. That is, placing the monomer on a 6-vertex implies a second monomer which is able to hop to the neighboring vertex via minimal monomer moves. Therefore, in a maximum matching the edge connecting these neighboring monomers will always be covered by a dimer. First, the presence of a monomer on the 6-vertex means that, by definition, no monomers can exist on neighboring vertices in a maximum matching. Consider the neighboring 3-vertex circled in black in Fig.~\ref{fig:even_loop_proof}a: this must have a dimer connected to one of the two edges not connected to the 6-vertex. The two choices are equivalent under a vertical mirror symmetry. Cover the green edge with a dimer as shown. Proceeding clockwise, a chain of dimers (purple) is implied in the numerical order indicated. If any of these is not covered by a dimer, an anti-monomer will result which connects via an augmenting path to the original monomer. At the end of this chain of implied dimers, the red monomer results. This monomer neighbors the original 6-vertex monomer, and provides the desired contradiction, as the number of dimers in the matching can always be increased by covering the edge connecting the two monomers. No monomer can ever appear on a 6-vertex, and, by extension, on any vertex connected by an alternating path to a 6-vertex, in a maximum matching. A similar argument demonstrates that 4-vertices also never host monomers in maximum matchings (see Appendix~\ref{app:even_loop_proof}, Fig.~\ref{fig:no_monomer_4} for a graphical proof). $\square$

A corollary of this result is that in a maximum matching, monomers cannot be placed on the $5_C$-vertices which connect even-valence vertices, except in a $4^5$ loop, where at most one $5_C$ vertex can be covered by a monomer. For a graphical proof, see Appendix~\ref{app:even_loop_proof}, \ref{fig:no_monomer_5c}. We have tested these results numerically in finite systems by generating maximum matchings using the Hopcroft-Karp algorithm and verifying that they satisfy these constraints. 

The above arguments constrain the placement of dimers in maximum matchings: \textbf{in any maximum matching, an even-valence vertex must always touch a dimer, which must always cover one of the edges indicated in Fig.~\ref{fig:even_loop_proof}b}. \emph{Proof}: the first part of the statement follows from the absence of monomers on even-valence vertices in maximum matchings. The second can almost be seen directly from the proof presented in Fig.~\ref{fig:even_loop_proof}a: the monomer placed on the 6-vertex could equally well be a dimer protruding downwards (any of the three options indicated as disallowed in Fig.~\ref{fig:even_loop_proof}b). However, there would be no immediate problem with a monomer residing next to such a dimer. To complete the proof, we only need to show that a monomer of the opposite bipartite charge always resides on the other side of the thick black line; the graphical proof of this is also presented in Appendix~\ref{app:even_loop_proof} (Figs.~\ref{fig:no_legs_6} and \ref{fig:no_legs_4}). Therefore, if present, the forbidden dimer would permit an augmenting path (which crosses the line). In a maximum matching, this situation can never arise, since by definition augmenting paths cannot exist. Rephrasing in terms of minimal monomer moves, the forbidden dimer would allow the passage of precisely one monomer over the line, which could then annihilate with a monomer of opposite charge. $\square$

\begin{figure*}[t]
\includegraphics[width=\textwidth]{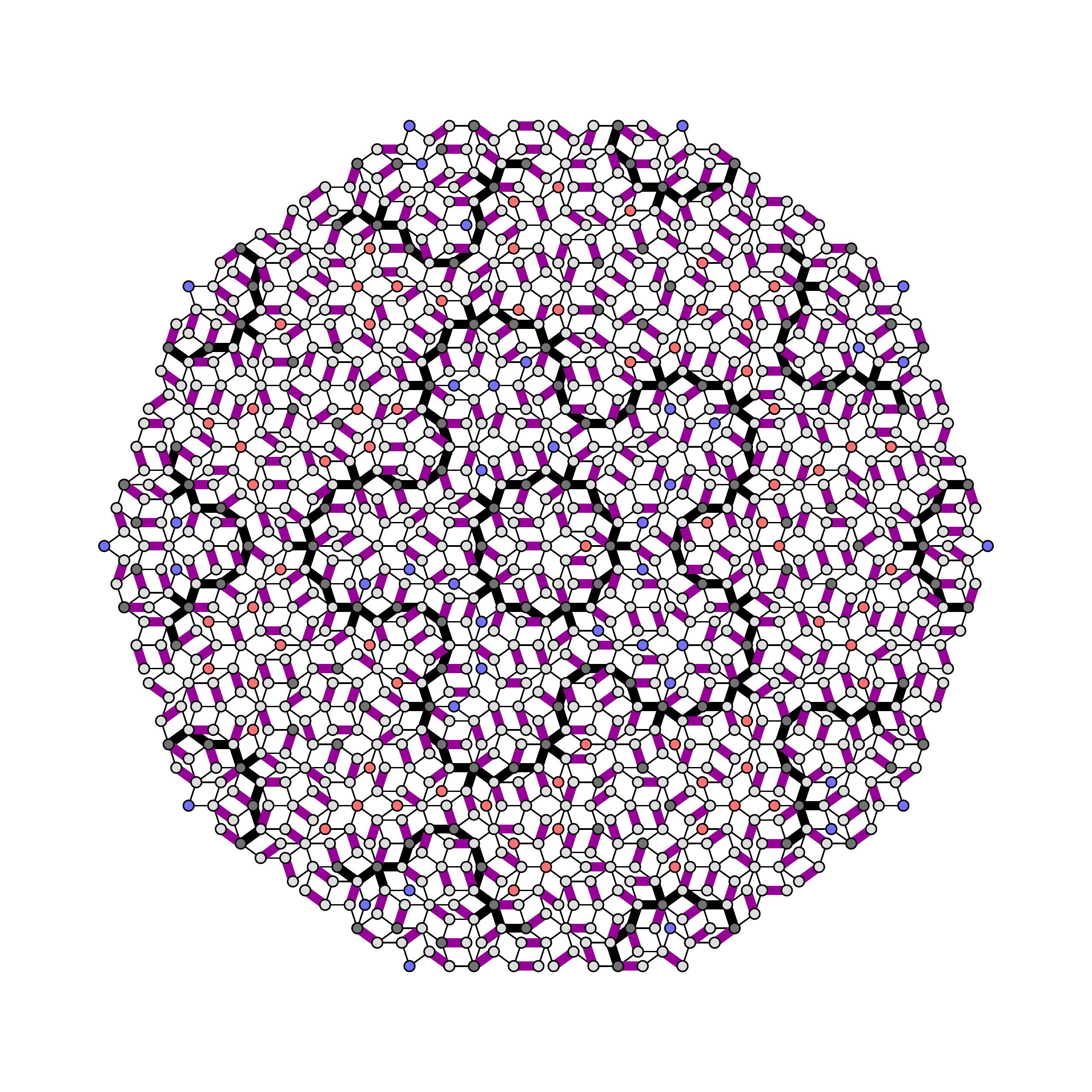}
\caption{A maximum matching of a finite section of the Penrose tiling. Certain edges (thick black lines) of even-valence vertices (dark grey) cannot be covered by dimers (purple) in maximum matchings. Monomers, colored blue or red according to their bipartite charge, cannot cross the closed thick black loops.
}
\label{fig:bigA}
\end{figure*}

Fig.~\ref{fig:bigA} shows a finite section of Penrose tiling. Thick black lines indicate the edges which can never be covered by dimers in a maximum matching. As expected, these form closed loops connecting second-nearest neighbor even-valence vertices, but also include twigs protruding from the loops at the 6-vertices. The twigs capture the added constraints on dimer placement on 6-vertices shown in Fig.~\ref{fig:even_loop_proof}b. We term these loops \emph{monomer membranes}, and their significance is as follows.
 
{\bf Monomer membranes provide impermeable barriers to monomer motion}. \emph{Proof}: each closed loop bounds two regions. On the side of the loop into which the even-valence vertices point (with their directions indicated in Fig.~\ref{fig:even_loop_proof}b), a dimer will protrude from each even-valence vertex. In principle this dimer could provide one end of an alternating path, the other end of which terminates on a monomer. However, the monomers in this region are all of the opposite bipartite charge to the even-valence vertices comprising the membrane, so this cannot occur. On the side of the loop away from which the even-valence vertices point, there exist monomers of the same bipartite charge as the even-valence vertices --- but no dimers connect from the even-valence vertices into this region, and, since alternating paths start on monomers and end on the dimer connecting to a vertex, no alternating path can reach these vertices. The situation is reversed for the $5_C$-vertices constituting the remaining members of the loops, which have opposite bipartite charge to the even-valence vertices. $\square$

\begin{figure*}[t]
\includegraphics[width=.99\textwidth]{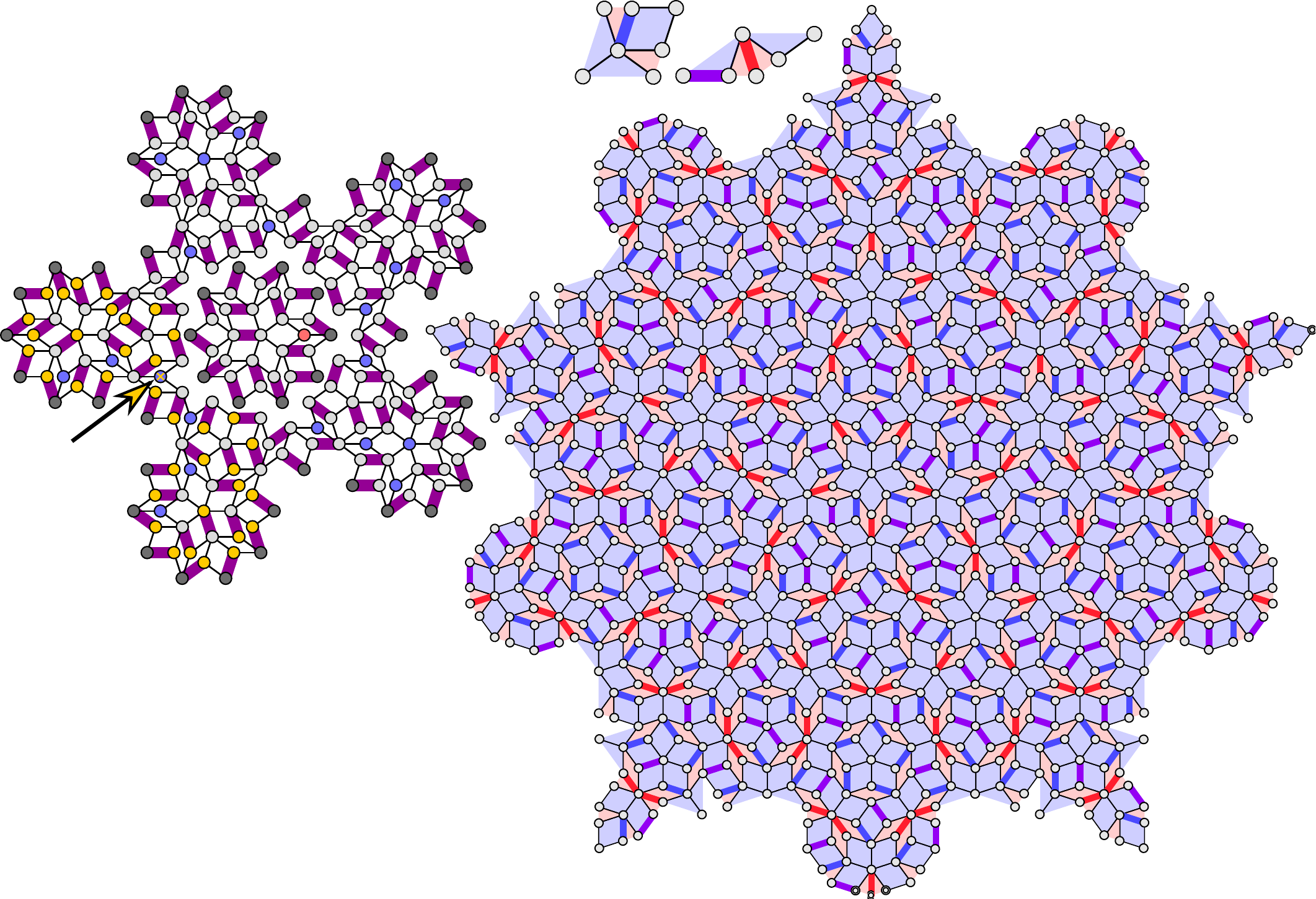}
\caption{$(a)$ Deleting uncoverable edges in Fig.~\ref{fig:bigA} disconnects the graph into \emph{monomer regions} (the incomplete outermost region has been removed). Neighboring monomer regions contain opposite net bipartite charge. The monomer indicated with a gold cross and arrow is able to reach the gold vertices by augmenting its alternating paths. It is not able to reach all vertices, even of its own charge; the obstruction is made by dimers rather than monomers (see Section~\ref{sec:Ergodicity}). $(b)$ Decorating the basic inflation elements with dimers in the final inflation results in the dimer covering shown. The following \emph{dimer inflation algorithm} gives a maximum matching: whenever a vertex is covered by two dimers, delete one. One monomer is then associated to each $5_{A,B}$-vertex with zero or one $7$-vertices as second-nearest neighbors, and three monomers are associated to each $5_A$-vertex with two $7$-vertices as second-nearest neighbors.
}
\label{fig:bigB}
\end{figure*}

The set of maximum matchings is unaffected if we delete the edges which can never be covered by dimers. As can be seen in Figs.~\ref{fig:bigA} and \ref{fig:bigB}, deleting these edges causes the graph to break into disconnected regions. Note that this illustrates the importance of the twigs --- without deleting these, the different regions remain connected. A careful inspection reveals that there is an edge of the $5_C$-vertex, when the $5_C$-vertex appears in the configuration $-4-6-4-\boxed{5_C}-4-4-6-4-$ (where only the relevant $5_C$ vertex is listed), which must also be deleted to cause the regions to disconnect. We prove in Appendix~\ref{app:even_loop_proof} that the edge is indeed uncoverable by a dimer in maximum matchings precisely whenever it appears in this configuration. Each disconnected region is a sub-graph containing an excess of one or the other bipartite charge. This forces a finite number of monomers of the corresponding excess charge in each region. Any two neighboring regions contain oppositely-charged monomers. The edges of the $4^5$ loop can be covered by dimers, and so does not result in disconnection.

The graphical proofs in this section and in Appendix~\ref{app:even_loop_proof} rely upon chains of dimers being implied by the placement of only two initial objects (a monomer and a dimer, or two dimers). These \emph{implication networks}, as we term them, always exist, and always close: once a dimer has been implied, a whole chain must be implied, terminating only once the resulting network returns to the original dimer (we allow dimers to continue to be implied after a monomer has been implied). The proof follows from two observations: (i)~constraining two legs of a 3-vertex not to host dimers
automatically implies a dimer on the third leg; and (ii)~at least two of the three vertices diametrically opposite a 3-vertex (\emph{i.e.}~across the three rhombuses that meet at the vertex) are themselves 3-vertices. The monomer membranes contain no 3-vertices; they form the boundaries to the implication networks~\footnote{We note that python's networkX implementation of the Hopcroft-Karp algorithm never seems to place monomers on the branching points of implication networks, despite the fact that the monomers it does place connect via alternating paths to these points. See for example Fig.~\ref{fig:bigA}.}.

\subsection{Monomer Membrane Properties}
\label{subsec:monomer_membrane_properties}

As the \emph{monomer membranes} (closed loops of second-nearest neighbor even-valence vertices connected via $5_C$ vertices) are central to understanding the maximum matchings of the Penrose tiling, we provide further details of their properties here.

\begin{figure}[t]
\includegraphics[width=.4\textwidth]{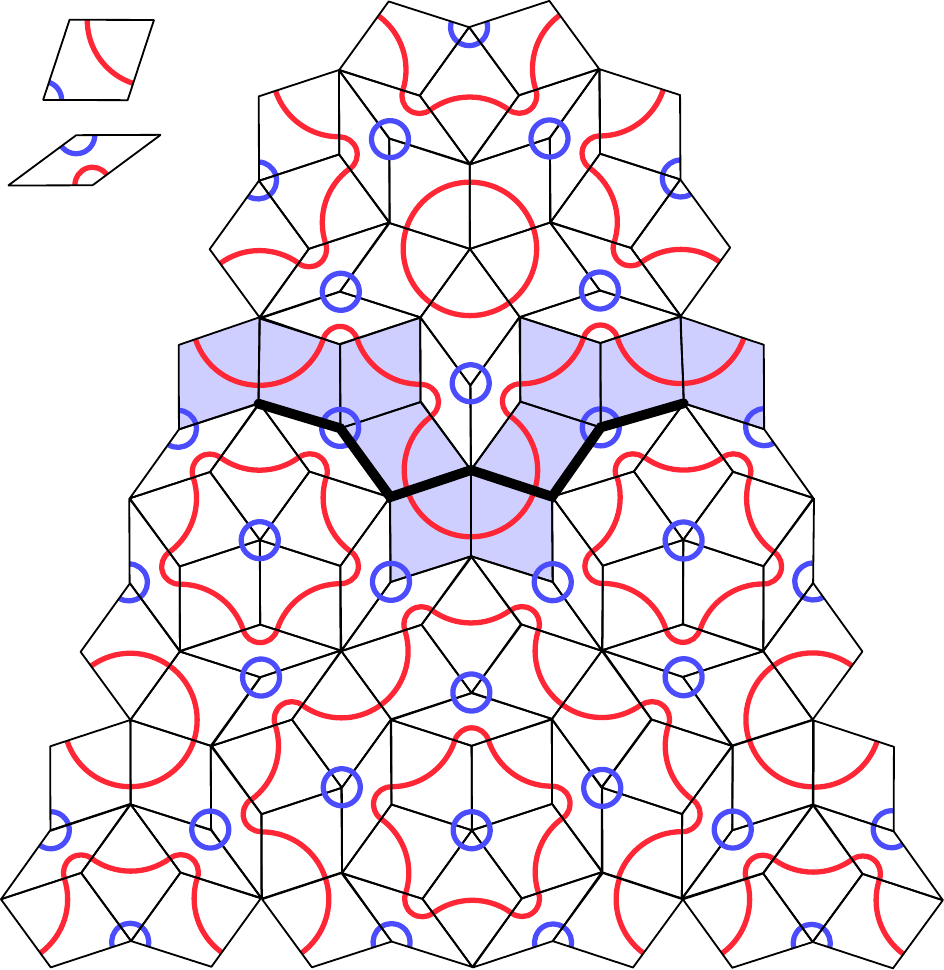}
\caption{Decorating the tiles with red and blue curves as indicated, continuity of the curves implies the matching rules of Fig.~\ref{fig:tiles}a. At most two red curves may cross the entire system~\cite{Gardner,GrunbaumShephard}. All others must close around regions of $D_5$ symmetry. The curves follow chains of thick rhombuses, one of which is highlighted in blue (all thick rhombuses are adjacent to precisely two others). The curves also follow chains of second-nearest-neighbor even-valence vertices (monomer membranes): a 4-6-4 segment is highlighted with a thick black line. Therefore these, too, close around regions of $D_5$ symmetry, centred on either a $5_A$- or $5_B$-vertex. The thick black line can be seen to curve away from the 4-vertex with curvature $2\pi/5$, and towards the 6-vertex with curvature $-4\pi/5$, with the directions of the vertices defined in Fig.~\ref{fig:even_loop_proof}.
}
\label{fig:loops}
\end{figure}

Fig.~\ref{fig:loops} shows an alternative decoration of the two rhombuses, with red and blue curves. Continuity of the curves enforces the tile matching rules (Fig.~\ref{fig:tiles}). Conway and Penrose independently demonstrated that at most two of the red curves may cross the entire tiling; all others form closed loops with $D_5$ symmetry ($5m$ in Hermann-Mauguin notation), which enclose regions of $D_5$ symmetry, each centred on either a $5_A$- or $5_B$-vertex~\cite{Gardner,GrunbaumShephard}. Every thick rhombus in the tiling is adjacent to precisely two other thick rhombuses, and therefore chains of thick rhombuses also either cross the system or form closed curves. The red curves can be seen to follow the thick-rhombus chains. Monomer membranes also follow thick-rhombus chains, crossing the chains twice at each 6-vertex, and staying on the same side of the chains around 4-vertices. The proof again follows from the fact that the local empire of the 6-vertex, shown in Fig.~\ref{fig:loops}, can cover the entire tiling, allowing for overlaps. The results of Conway and Penrose therefore carry over to monomer membranes: at most two monomer membranes cross the entire system. All others are closed, with $D_5$ symmetry, and bound $D_5$-symmetric sets of tiles. The proof that closed monomer membranes feature $D_5$ symmetry follows from the fact that each is generated by repeated inflation of a $5_A$- or $5_B$-vertex, each of which has $D_5$ symmetry, and the fact that inflation about a $D_5$-symmetric point preserves the symmetry. As system-spanning loops cannot have $D_5$ symmetry, they cannot appear in any Penrose tilings created by inflating $D_5$-symmetric tile sets. As we  focus primarily on such tile sets, the system-spanning loops will play a limited role in the subsequent discussion.

The local empire of the $5_A$-vertex is bounded by a $4^5$-loop, the smallest monomer membrane. All other monomer membranes can be generated as inflations of this case~\footnote{The sequence of central vertices under repeated inflation is $5_A$, $5_B$, $5_A$ rotated through $2\pi/10$, $5_B$ rotated through $2\pi/10$, after which the sequence repeats. Note that an arbitrary vertex takes at most three inflations to reach the sequence}. Inflating the $4^5$-loop twice returns a (rotated) $4^5$-loop with some surrounding tiles. For subsequent inflations, the list of even-valence vertices in the outermost monomer membrane under inflation is:
\begin{align}
4^{5}\rightarrow\left(46\right)^{5}\rightarrow\left(4644\right)^{5}\rightarrow\left(46444646\right)^{5}\rightarrow\ldots
\label{eq:MM_inflation}
\end{align}
where only the even-valence vertices in the loop are listed. The length of the boundary after $m$ inflations, as measured by the number of even-valence vertices it contains, is $5\times2^{m}$. This number is odd for $m=0$ ($4^5$, the only membrane permeable to monomers), and even for all $m>0$. 

The specific sequences of 4s and 6s can be generated by the substitution rules:
\begin{align}
4 &\rightarrow 46\nonumber\\
6 &\rightarrow 44.
\label{eq:MM_inflation_rules}
\end{align}
If $n_{4,6}$ and $n'_{4,6}$ denote the number of 4-,6-vertices before and after inflation, we may assign a growth matrix as in Section~\ref{subsec:background_penrose}:
\begin{align}
\left(\begin{array}{c}
n'_4\\
n'_6
\end{array}\right)=
\left(\begin{array}{cc}
1 & 2\\
1 & 0
\end{array}\right)\left(\begin{array}{c}
n_4\\
n_6
\end{array}\right).
\end{align}
The largest eigenvalue of the growth matrix is two, confirming that the loop length doubles under inflation. Since the number is rational, the loops formed by an infinite number of inflations are not themselves quasilattices~\footnote{Note that these inflation rules also govern the itinerary of the period-doubling cascade into chaos in unimodal maps (see Ref.~\onlinecite{ZaporskiFlicker18} and references therein)}.

The total curvature of any loop must be $2\pi$. The membranes curve away from the 4-vertex with curvature $2\pi/5$ (since the $4^5$ loop closes), and towards the 6-vertex with curvature $-4\pi/5$ (since the $\left(46\right)^5$ loop also closes). This can be seen, for example, in Fig.~\ref{fig:loops}. Monomer membranes of increasing size come in two varieties: one centered on a $5_A$-vertex, and one centered on a $5_B$-vertex. The sequence of vertices appearing in each of these two varieties of membranes (Eq.~\eqref{eq:MM_inflation}) can be generated by the following rule: starting from a 4 (respectively 46), generate the next term in the sequence by inserting $\left(464\right)$ to the left of each symbol in the previous sequence. Considering just the minimal repeat unit (which appears five times in the full loop) this gives:
\begin{align}
4&\rightarrow \left(464\right)4\rightarrow \left(464\right)4\left(464\right)6\left(464\right)4\left(464\right)4\rightarrow\ldots\nonumber\\
46 &\rightarrow \left(464\right)4\left(464\right)6 \rightarrow\ldots
\end{align}
matching the sequence of Eq.~\eqref{eq:MM_inflation}, also generated by Eq.~\eqref{eq:MM_inflation_rules}. Since $\left(464\right)$ is curvature-free, this construction preserves the curvature of the seed. This guarantees a $\pm2\pi$ curvature for the loop. In principle there are other sequences compatible with $D_5$ symmetry and closed loops, such as $\left(46464\right)^5$. This case proves incompatible with the tile matching rules of Fig.~\ref{fig:tiles}: any $D_5$-symmetric region of a Penrose tiling must be centred on a $5_A$- or $5_B$-vertex, but this sequence is centred on a \emph{decapod defect}~\cite{Gardner,GrunbaumShephard,BoyleSteinhardt16B}.

The monomer membranes are fractal objects, and are characterized by a fractal dimension $d_F$. We may define $d_F$ as follows: upon rescaling the area $\mathcal{A}$ via
\begin{align}
\mathcal{A}&\rightarrow\mathcal{A}\cdot\epsilon
\end{align}
a geometric quantity $\mathcal{S}_{d_F}$ of dimension $d_F$ scales to
\begin{align}
\mathcal{S}_{d_F} &\rightarrow \mathcal{S}_{d_F}\cdot\epsilon^{d_F/2}.
\end{align}
The average co-ordination of tiles in the Penrose tiling is four, meaning the number of vertices is equal to the number of tiles; thus under an inflation, the total number of vertices in the infinite tiling (hence the effective area) increases by a factor $\epsilon = \varphi^2$, whereas the number of vertices along a monomer membrane doubles. Hence, the fractal dimension of the monomer membrane is determined by setting $\left(\varphi^{2}\right)^{d_F/2}=2$, yielding
\begin{align}
d_F=\frac{1}{\log_{2}\varphi}\approx1.440.
\end{align}

%
\section{Maximum Matchings}
\label{sec:maximum_matchings}
%

In Section~\ref{sec:no_perfect_matchings} we established that Penrose tilings do not admit perfect matchings. Consequently, any dimer covering must necessarily include monomers. In this section, we proceed to find a set of maximum dimer matchings, which host the fewest possible monomers. We do so in two steps: first, in Section~\ref{subsec:dimer_inflation} we provide an algorithm for generating dimer matchings, and then in Section~\ref{subsec:dimer_inflation_proof} we prove that the matchings generated by this algorithm indeed contain the maximum number of dimers.

\subsection{Dimer Inflation Algorithm}
\label{subsec:dimer_inflation}

Recall that the Penrose tiling can be constructed by an inflation procedure. It is natural to ask whether we can leverage this to construct dimer coverings in a similar fashion. Consider a finite section of the Penrose tiling, generated by inflating a simply-connected tile set a finite number of times. We can imagine decorating the seed by placing dimers on certain edges before inflation. Strictly, the procedure to achieve $n$ inflations would be to inflate $n-1$ times via the standard inflation rules, then once with the decorated inflation rules which specify the dimer positions. As an example, Fig.~\ref{fig:bigB}b shows a finite section of the Penrose tiling obtained by performing four inflations of the local empire of the $5_A$-vertex using the standard inflation rules of Fig.~\ref{fig:tiles}, followed by a final inflation using the dimer-decorated inflation rules specified in Fig.~\ref{fig:bigB}b. However, this figure also illustrates a general obstruction to this procedure: namely, that it always leads to some vertices in the final covering which are covered twice by dimers (which is forbidden).

{\bf It is impossible to create a maximum matching with the full inflation symmetry of the tiles}. \emph{Proof}: this is seen most easily by observing that the three edges on the central red triangles of the thin tile become the three edges of the 6-vertex, one of which must be covered by a dimer in a maximum matching (Fig.~\ref{fig:even_loop_proof}b). Placing a dimer on any one of these edges leads to a forbidden double cover elsewhere in the inflated pattern. $\square$

However, a simple algorithm can be applied to any forbidden dimer covering (such as that in Fig.~\ref{fig:bigB}b) to produce an allowed matching. Whenever a vertex is covered by two dimers, simply delete one. This will create a monomer neighboring the formerly double-covered vertex. Note also that monomers existed already (generated by the inflation, not the deletion) at $5_B$-vertices and some $5_A$-vertices. In the case of the $5_B$ this is unavoidable, as all five of the edges meeting at these vertices come from the same edge in the inflation tiles (the edge lying on the mirror axis of the thick rhombus). 

\begin{figure}[t]
\includegraphics[width=.4\textwidth]{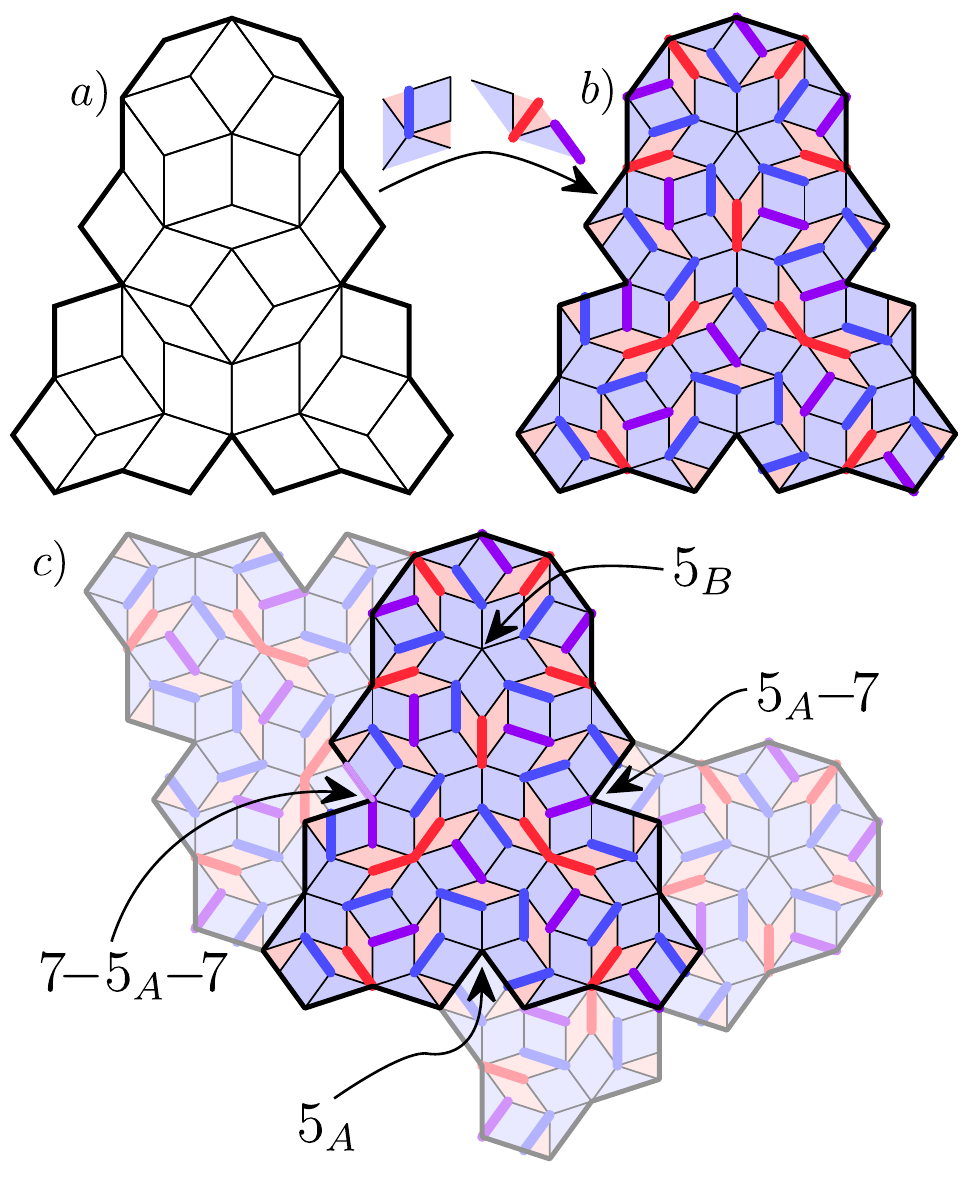}
\caption{
Proof that the dimer inflation algorithm of Section~\ref{subsec:dimer_inflation} associates monomers with $5_{A,B}$-vertices based on how many 7-vertices they have as second-nearest neighbors (points (i)-(iii) in that section). $(a)$ The local empire of the 4-vertex is large enough to cover the Penrose tiling, allowing for overlaps. $(b)$ Inflating using the dimer-decorated inflation rules leads to a set of tiles large enough to cover the inflated tiling. $(c)$ To confirm statements (i)-(iii) about monomer placement, it is necessary to check every continuation of the boundary vertices. Two continuations are shown, with relevant features highlighted: a $5_B$-vertex, and $5_A$-vertices with zero, one, and two 7-vertices as second-nearest neighbors.
}
\label{fig:DIA_proof}
\end{figure}

This construction, which we term the {\it dimer inflation algorithm}, can be seen to place monomers as follows:
\begin{enumerate}[label=(\roman*),itemsep=0pt]
\item if a $5_A$- or $5_B$-vertex has no 7-vertices as second-nearest neighbors, a monomer is placed with the same bipartite charge as the 5-vertex
\item if a $5_A$-vertex has one 7-vertex as a second-nearest neighbor, a monomer is placed with opposite bipartite charge to the 5-vertex
\item if a $5_A$ vertex has two 7-vertices as second-nearest neighbors, three monomers are placed with opposite bipartite charge to the 5-vertex.
\end{enumerate}
{\bf These are the only monomers placed by the Dimer Inflation Algorithm.} \emph{Proof:} allowing for overlaps, the local empire of the 4-vertex is sufficiently large that it can cover the entire Penrose tiling. Therefore, the inflation of the local empire of the 4-vertex can cover the inflated tiling. Fig.~\ref{fig:DIA_proof}a shows the local empire of the 4-vertex, and Fig.~\ref{fig:DIA_proof}b its inflation by the dimer-decorated inflation tiles. This tile set, including its dimer decoration, covers the entire tiling, and it suffices to check that the dimer inflation algorithm only places monomers according to (i)-(iii). A $5_B$-vertex appears in the tile set, and hosts a monomer, in agreement with (i). All $5_A$-vertices appear on the boundary of the tile set upon different continuations. All possible continuations of the boundary, consistent with the matching rules, must be checked. Fig.~\ref{fig:DIA_proof}c shows two continuations of the tile set; the combined tile set features all combinations of $5_{A,B}$- and 7-vertices, and all obey the stated monomer placement rules. We have checked all possible boundary completions and confirmed that statements (i)-(iii) hold, and that no further monomers are created by the dimer inflation algorithm. $\square$

As there are 12 edges in total between the two once-inflated tiles, there are $2^{12}$ choices for dimer coverings consistent with inflation. We prove in Section~\ref{subsec:dimer_inflation_proof} that the algorithm considered here leads to a maximum dimer matching, and so no other choice can do better. However, we found several other examples which do as well as this choice, albeit with slightly more complicated rules for removing double dimer coverings of vertices. 

\subsection{The Dimer Inflation Algorithm Produces Maximum Matchings}
\label{subsec:dimer_inflation_proof}

\begin{figure}[t]
\includegraphics[width=.4\textwidth]{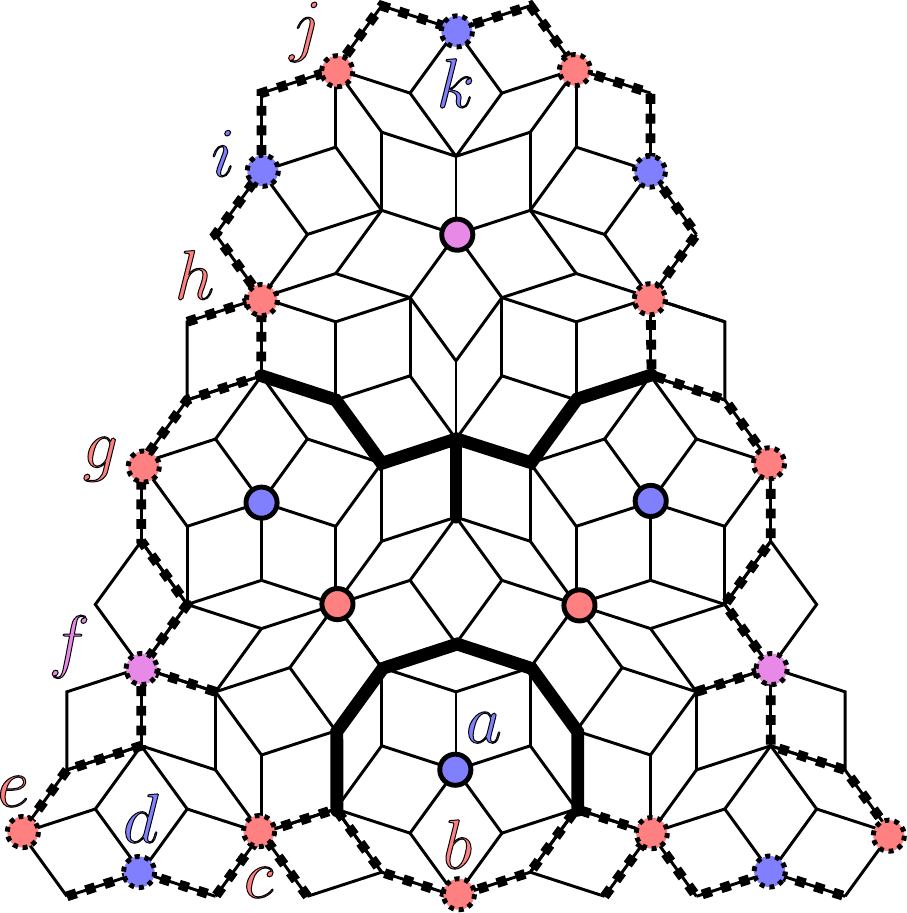}
\caption{The local empire of the $6$-vertex with key vertices identified, used in the proof in Section~\ref{subsec:dimer_inflation_proof} and Appendix~\ref{app:dimer_inflation_proof} that the dimer inflation algorithm in Fig.~\ref{fig:bigB}b generates maximum matchings. Solid (dashed) circles indicate vertices of definite (potential) valence: red 7-vertices, blue $5_A$-vertices, and pink $5_B$-vertices. Solid (dashed) thick black lines indicate definite (potential) segments of monomer membranes (which cannot be covered by dimers or crossed by monomers).
}
\label{fig:algorithm_maximum_proof}
\end{figure}

To see that the matching generated by the dimer inflation algorithm in Fig.~\ref{fig:bigB}b is maximum, recall the following facts. First, the local empire of the 6-vertex (Fig.~\ref{fig:algorithm_maximum_proof}) is sufficiently large that it can cover the entire Penrose tiling, allowing for overlaps~\cite{Gummelt96}. Second, monomers cannot cross monomer membranes (solid thick black edges in Fig.~\ref{fig:algorithm_maximum_proof} for even-loop segments, dashed thick black edges for potential even-loop segments). Third, the dimer inflation algorithm creates monomers as stated in points (i)--(iii) of Section~\ref{subsec:dimer_inflation}. If we can prove that no monomers placed by the algorithm are connected by augmenting paths, then the matching must be maximum. The only obstacles to augmenting paths are monomer membranes. We only need to check the relationships between the 5- and 7-vertices, rather than the monomers themselves, as the relationship between the monomers and the vertices is specified in (i)--(iii). 

For example, any path connecting any two $5_B$-vertices must contain an even number of edges (be of even length) if it does not cross an impermeable monomer membrane. This is because each $5_B$-vertex has associated with it a monomer of the same bipartite charge in a maximum matching. Two of these monomers must have the same charge if they are not separated by a monomer membrane. If they had opposite charge they would be connected by an augmenting path, and the matching would not be maximum. Therefore any two $5_B$-vertices connected by a path not crossing a monomer membrane must also have the same bipartite charge. As a corollary, any path connecting two $5_B$-vertices and not crossing a monomer membrane must itself be of even length.

In fact, not only does this result hold, but so does a stronger one: any path connecting any two $5_B$-vertices crosses an even (odd) number of edges if it crosses impermeable monomer membranes an even (odd) number of times. This implies not only that the $5_B$-related monomers within one region are of the same bipartite charge, but those within neighboring regions are of opposite bipartite charge. Consider the local empire of the 6-vertex in Fig.~\ref{fig:algorithm_maximum_proof}. It contains one certain $5_B$-vertex (pink, solid circle) in the bulk, and two potential $5_B$-vertices on the boundary related by a mirror symmetry (pink, dashed circles). The solid thick black lines indicate segments of the impermeable monomer membrane, and the dashed thick black lines indicate potential monomer membrane segments. As the membrane forms a closed loop, it must separate the known $5_B$-vertex from the two potential $5_B$-vertices. We see that these vertices are separated by paths of odd length. The two boundary pink potential $5_B$-vertices are separated by paths of even length. However, if they are both $5_B$-vertices, this resolves the ambiguity of the potential monomer membranes passing through them, and confirms that they are not separated by a monomer membrane, which is correct. As the entire tiling can be constructed from this local empire, there can be no other possible relations between $5_B$-vertices. 

\begin{figure}[t]
\includegraphics[width=.3\textwidth]{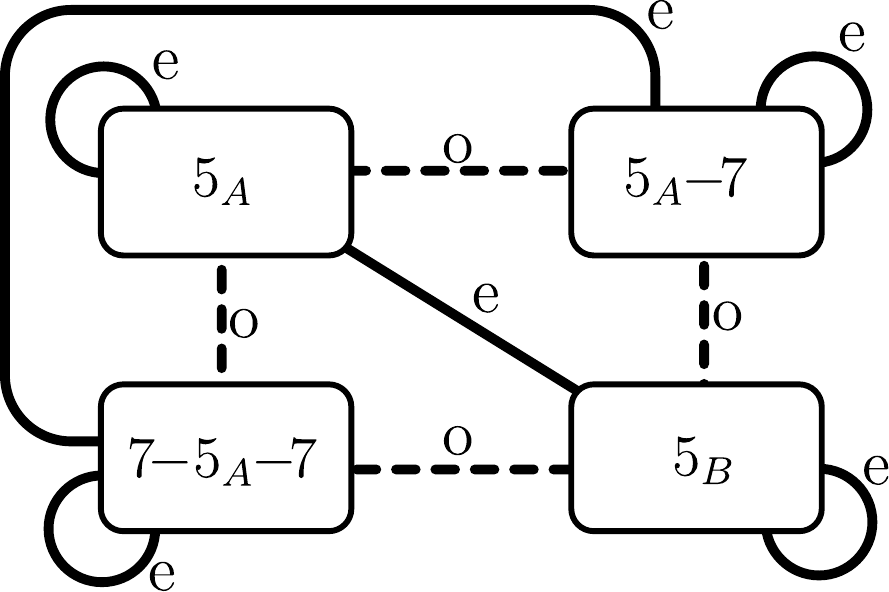}
\caption{The listed vertex types must be separated by paths of even (e, solid lines) or odd (o, dashed lines) length {if the vertices are separated by an even number of monomer membranes (including zero). If the vertices are separated by an odd number of monomer membranes, e and o should be interchanged}. The box labeled `$5_A$' excludes the `$5_A$--7' and `7--$5_A$--7' configurations, and `$5_A$--7' excludes `7--$5_A$--7'. See Fig.~\ref{fig:DIA_proof} and Section~\ref{subsec:dimer_inflation_proof}. 
}
\label{fig:connections}
\end{figure}

To complete the proof it suffices to prove the following statements:
\begin{itemize}[itemsep=0pt]
\item any path connecting any two 7-vertices is of even (odd) length if it crosses impermeable monomer membranes an even (odd) number of times
\item any path connecting any two $5_{A,B}$-vertices, where the $5_{A,B}$-vertices have no 7-vertices as second-nearest neighbors, is of even (odd) length if it crosses impermeable monomer membranes an even (odd) number of times
\item any path connecting any 7-vertex to any $5_{A,B}$-vertices, where the $5_{A,B}$-vertices have no 7-vertices as second-nearest neighbors, is of odd (even) length if it crosses impermeable monomer membranes an even (odd) number of times.
\end{itemize}
The statements are summarized in Fig.~\ref{fig:connections}. Note that the only monomer membrane permeable to monomers is the $4^5$ loop. The results are evident for the interior of the local empire of the 6-vertex; the only cases to check are therefore any (ambiguous) boundary vertices which have the potential to become one of the vertices in question.

The options are presented in Fig.~\ref{fig:algorithm_maximum_proof}. Red vertices are 7-vertices (solid circles indicate certain 7-vertices, whereas dashed circles indicate potential 7-vertices); blue vertices are (potential or actual) $5_A$-vertices; pink vertices are (potential or actual) $5_B$-vertices. The verification that every case works is lengthy but straightforward, and appears in Appendix~\ref{app:dimer_inflation_proof}. In all possible cases, the three listed requirements are fulfilled, and the dimer inflation algorithm given in Section~\ref{subsec:dimer_inflation} generates maximum matchings.

%
\section{Monomer Densities in Maximum Matchings}
\label{sec:monomer_densities}
%

Having presented, in Section~\ref{sec:maximum_matchings}, an algorithm for generating maximum matchings of the Penrose tiling, we proceed to analytically calculate (in Section~\ref{subsec:analytic_density}) and numerically check (in Section~\ref{subsec:numerical_density}) the density of monomers in any maximum matching.

\subsection{Analytic Calculation of the Minimal Monomer Density}
\label{subsec:analytic_density}

The results of Section~\ref{sec:maximum_matchings} reveal that we can associate monomers with $5_A$-vertices, $5_B$-vertices, and 7-vertices. The monomers are not constrained to sit at specific locations relative to these vertices, as they can move to any vertices to which they connect via alternating paths. However, we are able to use the association between monomers and vertices implied by the dimer inflation algorithm (points (i)--(iii) in Section~\ref{subsec:dimer_inflation}) to establish the density of monomers in maximum matchings of the Penrose tiling.

Table \ref{tab:frequencies} lists the relative frequencies of each vertex type in the Penrose tiling~\cite{Henley86,Jaric86,Peyriere86}. If we consider not just the vertex and its inflation, but the local empires of each, certain facts become evident. A $5_{A,B}$-vertex with no 7-vertices as second-nearest neighbors is created by inflating a $5_{B,A}$-vertex; this pattern is associated with one monomer. A $5_A$-vertex with one 7-vertex as a second-nearest neighbor is created by inflating a 6-vertex; this pattern is also associated with one monomer. Finally, a $5_A$-vertex with two 7-vertices as second-nearest neighbors is created by inflating a 7-vertex; this pattern is associated with three monomers. New vertices are also generated between existing vertices upon inflation, but inspecting the inflated vertices in Fig.~\ref{fig:vertices} reveals that only $3_B$- and $5_C$-vertices appear in this way.

\begin{table}
\noindent \begin{centering}
\begin{tabular}{|c|c|c|c|c|}
\hline 
dart-kite & rhombus & inflation & frequency & simplified\tabularnewline
\hline 
\hline 
Queen & $3_{A}$ & 7 & $\varphi^{3}\left(\varphi^{4}-1\right)$ & $11\varphi+7$\tabularnewline
\hline 
Deuce & $3_{B}$ & $3_{A}$ & $\varphi^{5}\left(\varphi^{4}-1\right)$ & $29\varphi+18$\tabularnewline
\hline 
King & 4 & 6 & $\varphi^{2}\left(\varphi^{4}-1\right)$ & $7\varphi+4$\tabularnewline
\hline 
Star & $5_{A}$ & $5_{B}$ & $\varphi^{4}$ & $3\varphi+2$\tabularnewline
\hline 
Sun 5 & $5_{B}$ & $5_{A}$ & $\varphi^{2}$ & $\varphi+1$\tabularnewline
\hline 
Jack & $5_{C}$ & 4 & $\varphi^{4}\left(\varphi^{4}-1\right)$ & $18\varphi+11$\tabularnewline
\hline 
Sun 4 & 6 & $5_{A}-7$ & $\varphi^{4}-1$ & $3\varphi+1$\tabularnewline
\hline 
Sun 3 & 7 & $7-5_{A}-7$ & $\varphi\left(\varphi^{4}-1\right)$ & $4\varphi+3$\tabularnewline
\hline
Ace & --- & --- & $\varphi^{6}\left(\varphi^{4}-1\right)$ & $47\varphi+29$\tabularnewline
\hline 
\end{tabular}
\par\end{centering}
\caption{\label{tab:frequencies}
The vertices in the dart-kite P2 tiling (Fig.~\ref{fig:vertices}); their equivalents in the rhombic P3 tiling; their inflations in P3; their relative frequencies; simplified forms of the frequencies, making use of the defining equation of the golden ratio $\varphi^2=\varphi+1$.}
\end{table}

Putting the results together, we find that the density of monomers $\rho_{\text{monomer}}$ in the tiling is
\begin{align}
\rho_{\text{monomer}}=\frac{f\left(5_A\right)+f\left(5_B\right)+f\left(6\right)+3f\left(7\right)}{\varphi^2\sum_{v\in\textrm{vertices}} f\left(v\right)}
\end{align}
where $f\left(v\right)$ is the relative frequency with which a $v$-vertex appears (see Table \ref{tab:frequencies}). The factor of $\varphi^2$ in the denominator arises because it is the inflated pattern which is to be compared to, and the number of all vertices increases by $\varphi^2$ under inflation in the infinite tiling. Substituting the values from Table \ref{tab:frequencies} we find the result for the monomer density:
\begin{align}
\rho_{\text{monomer}}=81-50\varphi
\end{align}
where the simplified result follows from repeated use of the defining equation of the golden ratio, $\varphi^2=\varphi+1$.

\subsection{Numerical Confirmation of the Minimal Monomer Density}
\label{subsec:numerical_density}

In order to confirm the analytic result of the previous section we carried out numerical calculations of the monomer densities in finite size sections of the Penrose tiling. To generate the sections we carried out up to thirteen inflations on seed tile configurations using the inflation rules of Fig.~\ref{fig:tiles}. We used as the seeds only the two basic rhombuses, thick and thin. Any other seed would simply be a combination of these, differing only on the boundary.

\begin{figure}[t]
\includegraphics[width=.4\textwidth]{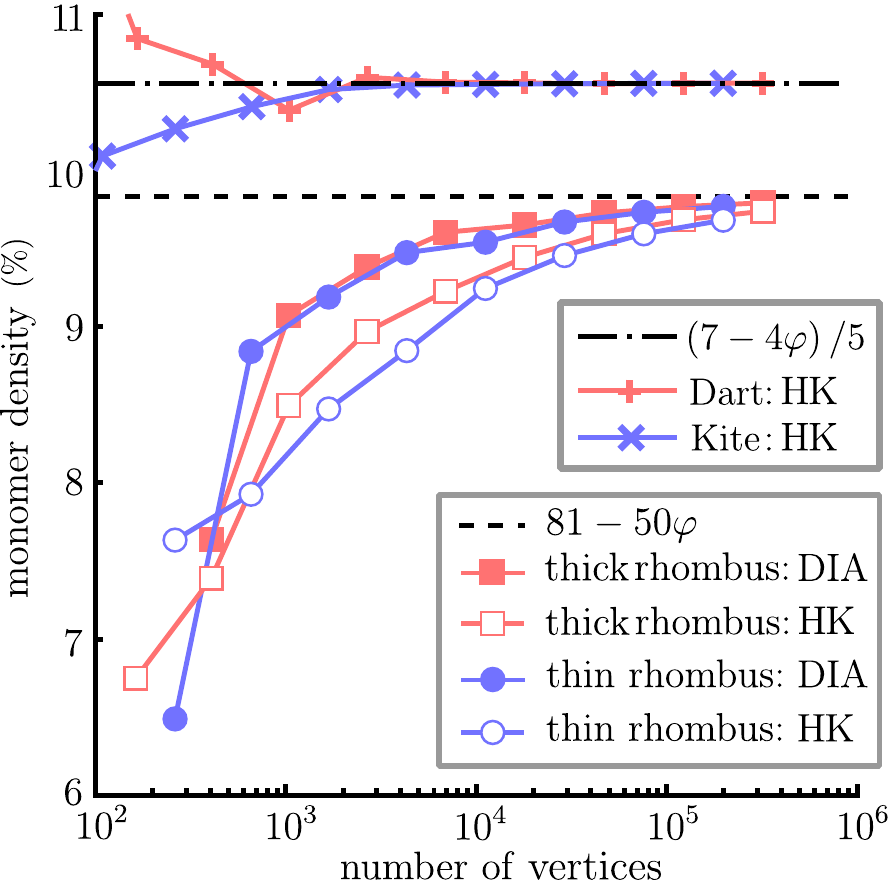}
\caption{Numerical results for the density of monomers in finite sections of the Penrose tiling. To achieve results with the least possible bias we consider inflations of each basic tile: the thick (red squares) and thin (blue circles) rhombuses for P3, and the dart (red crosses) and kite (blue crosses) for P2 (see Section~\ref{sec:other_quasilattices}). For the rhombic tiling, solid symbols correspond to systems in which monomers are placed according to the dimer inflation algorithm (DIA) of Section~\ref{subsec:dimer_inflation}, whereas hollow symbols correspond to systems in which maximum matchings are found using the Hopcroft-Karp algorithm (HK, also used for the dart-kite tiling). The analytical results, of $81-50\varphi\approx 0.098$ for the rhombic P3 tiling and $\left(7-4\varphi\right)/5\approx 0.106$ for the dart-kite P2 tiling, are shown by the black dashed lines.
}
\label{fig:numerical_density}
\end{figure}

In Fig.~\ref{fig:numerical_density} we show the numerically-obtained monomer densities (number of monomers in the finite system divided by the total number of vertices in the system). The inflation method can leave stray twigs on the boundary of the system; we prune these before finding maximum matchings. We consider two possible methods of placing monomers: the dimer inflation algorithm outlined in Section~\ref{subsec:dimer_inflation} (solid red squares for inflations of the thick tile, solid blue circles for inflations of the thin tile), and the Hopcroft-Karp algorithm~\cite{HopcroftKarp73}. We tested the Hopcroft-Karp result against an alternative maximum matching method, the Eppstein algorithm, and obtained identical results.

Differences between the numerical and analytical results may be attributed both to the usual boundary effects in a finite system, and to the impossibility of reproducing an irrational number as a ratio of two integers. In Fig.~\ref{fig:numerical_density} we see that both methods of monomer placement lead to monomer densities tending rapidly to the analytic value (note the logarithmic scale on the horizontal axis). The dimer inflation algorithm is less affected by the boundaries than the Hopcroft-Karp matching. The dimer inflation algorithm incurs errors in monomer placement only within one vertex from the boundary, when the boundary artificially decreases the valence of the vertex. The Hopcroft-Karp method, on the other hand, can take advantage of augmenting paths connecting bulk monomers with monomers artificially created by the boundary, and these paths can be of arbitrary length provided they are not constrained by monomer membranes. 

%
\section{The Maximum Matching Manifold is Connected by Minimal Monomer Moves}
\label{sec:Ergodicity}
%

In this section we demonstrate that, starting from one maximum matching, all others can be reached by minimal monomer moves without passing through non-maximum matchings. The set of maximum matchings therefore forms a connected manifold. 

To motivate this idea, consider a general bipartite graph which admits a perfect matching. Deleting a dimer from such a matching would create two neighboring monomers of opposite bipartite charge. In physical models, such a pair-creation process would be expected to be energetically costly. Assigning an energy to pair creation, the set of perfect matchings would constitute the degenerate set of classical ground states of the system. Starting from one perfect matching, if all others can be accessed by reconfiguring dimers in a physical manner (discussed shortly) while remaining within the set of perfect matchings, the set of degenerate classical ground states would form a connected manifold~\cite{MoessnerChalker98}. This would be a necessary but not sufficient condition for the system to be \emph{ergodic}, meaning that the time spent under dynamical evolution in a given volume of phase space of equal-energy microstates is proportional to the volume itself, or, equivalently, that all accessible microstates are sampled equally over sufficiently-long timescales~\cite{Boltzmann}. In the present work we are not specialising to a particular energy function or physical model, and the results can be considered to be in the zero-temperature limit. 

The sort of features which could disconnect a set of matchings might be, for example, the presence of non-contractible system-spanning loops under periodic boundary conditions. These can disconnect the phase space into topological sectors unreachable from one another by local (\emph{i.e.}~physical) moves~\cite{MoessnerRaman07}. Before considering the case of the Penrose tiling, it is instructive to continue to consider an arbitrary bipartite graph admitting multiple perfect matchings. In the case that only one topological sector exists, all perfect matchings can be enumerated by the following process: starting from one perfect matching (which could be found, for instance, by the Hopcroft-Karp algorithm), another can be generated by augmenting an alternating cycle. In fact, all perfect matchings can be enumerated by augmenting all alternating cycles in a given perfect matching~\cite{Uno97}. The following physical analogy motivates such moves: deleting one dimer on the cycle would create a monomer-antimonomer pair. The monomer can then hop around the cycle via minimal monomer moves, and re-annihilate with the anti-monomer, with the net effect being the desired augmentation. The analogy is imperfect, however, as the intermediate stages are not contained within the set of perfect matchings. In quantum dimer models they could be considered virtual processes.

Now consider the case of a finite bipartite graph which does not admit a perfect matching. Starting from a maximum matching, we further restrict to the case that every alternating cycle connects to at least one monomer via alternating paths. All maximum matchings can now be enumerated by identifying and augmenting all alternating cycles, plus all alternating paths (by definition each alternating path terminates on precisely one monomer)~\cite{Uno97}. The physical analogy holds more closely in this case. Augmenting an alternating path can be achieved simply by hopping the monomer along the path with minimal monomer moves. Augmenting an alternating cycle can be achieved by hopping a monomer to any vertex on the cycle, hopping the monomer around the cycle, then hopping the monomer back to its initial vertex along the same path it took to the cycle. In this process the only change to the dimer configuration is to augment the alternating cycle, since dimers on the path connecting the starting vertex to the cycle are returned to their initial configuration when the monomer retraces its steps.

Since no pair-creation occurs, these processes stay within the set of maximum matchings. The restriction that all alternating cycles connect via alternating paths to monomers means that it is always possible to find a monomer to carry out such a move. A simple example of a graph violating this condition would be to add to this example a second graph, disconnected from it, which itself admits multiple perfect matchings: the total graph therefore still does not admit a perfect matching, but the perfectly-matched region still requires the unphysical virtual processes to augment its alternating cycles.

Recall that the Penrose tiling divides into \emph{monomer regions} bounded by monomer membranes. At most two monomer membranes (\emph{i.e.}~one monomer region) can cross the entire system. The others are closed, with $D_5$ symmetry. Since monomer membranes can never host dimers in maximum matchings, we can consider each monomer region separately. {\bf Every monomer region contains an excess of bipartite charge}. \emph{Proof}: this can be seen (admittedly in a slightly roundabout way) from the fact that (i) the dimer inflation algorithm generates maximum matchings with monomers on every $5_B$-vertex, (ii) every monomer region contains a $5_B$-vertex (since the local empire of the 6-vertex contains $5_B$-vertices on both sides of the impermeable membrane, and covers the entire tiling), and (iii) the total number of monomers in a monomer region is equal to the monomer region's net bipartite charge. $\square$

{\bf All vertices with the same bipartite charge as their monomer region connect via alternating paths to monomers}. \emph{Proof}: we just proved that every monomer region contains at least one monomer. To see that every vertex within a monomer region (where the vertex has the same bipartite charge as the region) connects to a monomer via an alternating path, consider the following argument. Pick an arbitrary vertex $v_0$ within the monomer region with the same charge as the monomers in that region. If the vertex hosts a monomer, we are done. If not, the vertex must be covered by a dimer. Consider the vertex at the other end of the dimer, $v_1$. All the other edges of $v_1$ must be uncovered. Consider these edges, and the vertices $v_2^i$ (where $i+1$ is the valence of $v_1$) to which they connect. If any of $v_2^i$ hosts a monomer, this monomer then connects via an alternating path $v_2^i-v_1-v_0$ to the original vertex $v_0$, and we are done. If not, each of $v_2^i$ is covered by precisely one dimer. By iterating this process, a monomer must eventually be reached. $\square$

Since the Penrose tiling is bipartite and divides into closed regions in which every alternating cycle connects via an alternating path to a monomer, the set of maximum matchings forms a manifold connected by minimal monomer moves.

It is not true that any monomer within a monomer region connects via an alternating path to any given vertex with the same bipartite charge as the region. In terms of minimal monomer moves, a re-arrangement of the remaining monomer positions may in general be necessary to facilitate a monomer reaching a given vertex. Fig.~\ref{fig:bigB}a shows two monomer regions of the Penrose tiling obtained from Fig.~\ref{fig:bigA} by removing edges which can never be covered by dimers in maximum matchings, then by removing the largest connected component of the resulting disconnected graph (the outermost monomer region in Fig.~\ref{fig:bigA}. One monomer is highlighted with an arrow and gold cross; the set of vertices connected to this monomer by alternating paths is highlighted in gold. This is the set of vertices the monomer can reach via minimal monomer moves, holding all other monomers fixed. The dark grey vertices are even-valence vertices before disconnection, and form the unreachable monomer membrane which bounds the monomer region. Other vertices intermingled with the gold vertices are of the wrong bipartite charge and also cannot be reached. However, there are many more vertices within the region which are legitimate sites for occupation, and simply cannot be reached. The obstacle is not directly provided by other monomers, although these can also in principle form obstructions; instead, it is the dimer network. By moving monomers other than the crossed monomer it is possible to re-arrange the dimer network to allow the crossed monomer to reach any vertex.

%
\section{Classical Dimers on Other Penrose-like Tilings}
\label{sec:other_quasilattices}
%

All rhombic Penrose tilings are locally isomorphic to one another. Different decorations of the tiles can lead to alternative Penrose tilings in different local isomorphism classes. As the resulting graph connectivity changes under such decorations, the properties of maximum matchings may also change. In order to place our results for the rhombic Penrose tiling in a more general context, in this section we consider dimer coverings of other two-dimensional Penrose-like tilings. 

\subsection{Other Penrose Tilings}
\label{subsec:other_Penrose}

The P3 rhombic Penrose tiling was the third to be identified. Earlier versions include the dart-kite tiling P2, and the original pentagonal tiling P1, which has four different tile types. Fig.~\ref{fig:tiles} shows decorations of the P3 tiles which leads to the P1 and P2 tilings~\cite{BaakeGrimm}. Since one of the P1 tiles is a pentagon, P1 necessarily contains cycles of odd length, and the corresponding graph is therefore not bipartite. For this reason, we do not discuss it further. In this section we consider the dimer-covering properties of the dart-kite P2 tiling, and provide another simple decoration of P3 which trivially results in a perfect matching. 

\subsubsection{The Penrose Dart-Kite (P2) Tiling}

\begin{figure}[t]
\includegraphics[width=.4\textwidth]{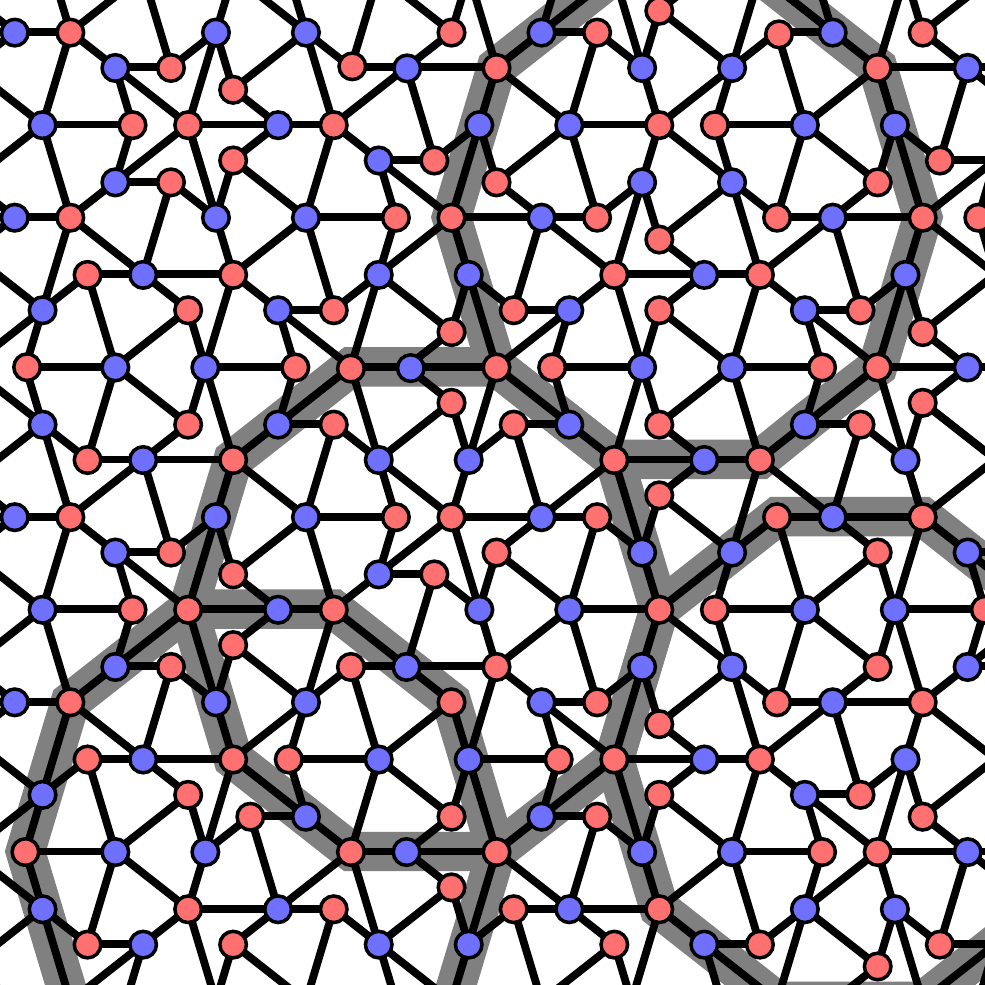}
\caption{A region of the Penrose dart-kite (P2) tiling. The vertices (Fig.~\ref{fig:vertices}) divide into bipartite sub-lattices: Star-Queen-King-Ace (red) and Deuce-Jack-Suns (blue). Dark grey decagons indicate \emph{cartwheels}, regions which can cover the infinite tiling allowing for overlaps~\cite{Gummelt96}.
}
\label{fig:P2}
\end{figure}

The Penrose P2 tiling consists of two tiles, referred to by Penrose as the \emph{dart} and \emph{kite}. Fig.~\ref{fig:tiles} shows how to derive a P2 tiling by decorating the rhombic tiles of P3. Alternatively,  P2 can be obtained by its own inflation rules. As both tiles have four sides, the graph is again bipartite. A finite region of the P2 tiling is shown in Fig.~\ref{fig:P2}. P2 features nine vertex types. Following \onlinecite{Henley86} we term them the Deuce, Jack, Queen, King, Ace, Star, and Suns 3--5 (as with $5_A$ and $5_B$ in the rhombic tiling, the different Sun vertices are distinguished by their neighboring tiles; the number indicates the number of dart tiles pointing out from the Sun). Except the Ace, all vertices are in one-to-one correspondence with vertices in the rhombic tiling; the correspondence is given in Fig.~\ref{fig:vertices} and Table~\ref{tab:frequencies}. Inspecting the local empires of each vertex reveals an interesting property: each vertex belongs to precisely one bipartite sub-lattice. The Suns, Deuce, and Jack form the entirety of one sub-lattice (blue vertices in Fig.~\ref{fig:P2}, and the Star, Queen, King, and Ace form the other (red). This is in contrast to P3 in which each vertex appears in both sub-lattices. The P2 tiling admits a covering (allowing for overlaps) by a set of tiles known as a \emph{cartwheel}~\cite{Gardner,Gummelt96,BaakeGrimm}. Some cartwheels are highlighted in Fig.~\ref{fig:P2}.

\begin{figure}[t]
\includegraphics[width=.4\textwidth]{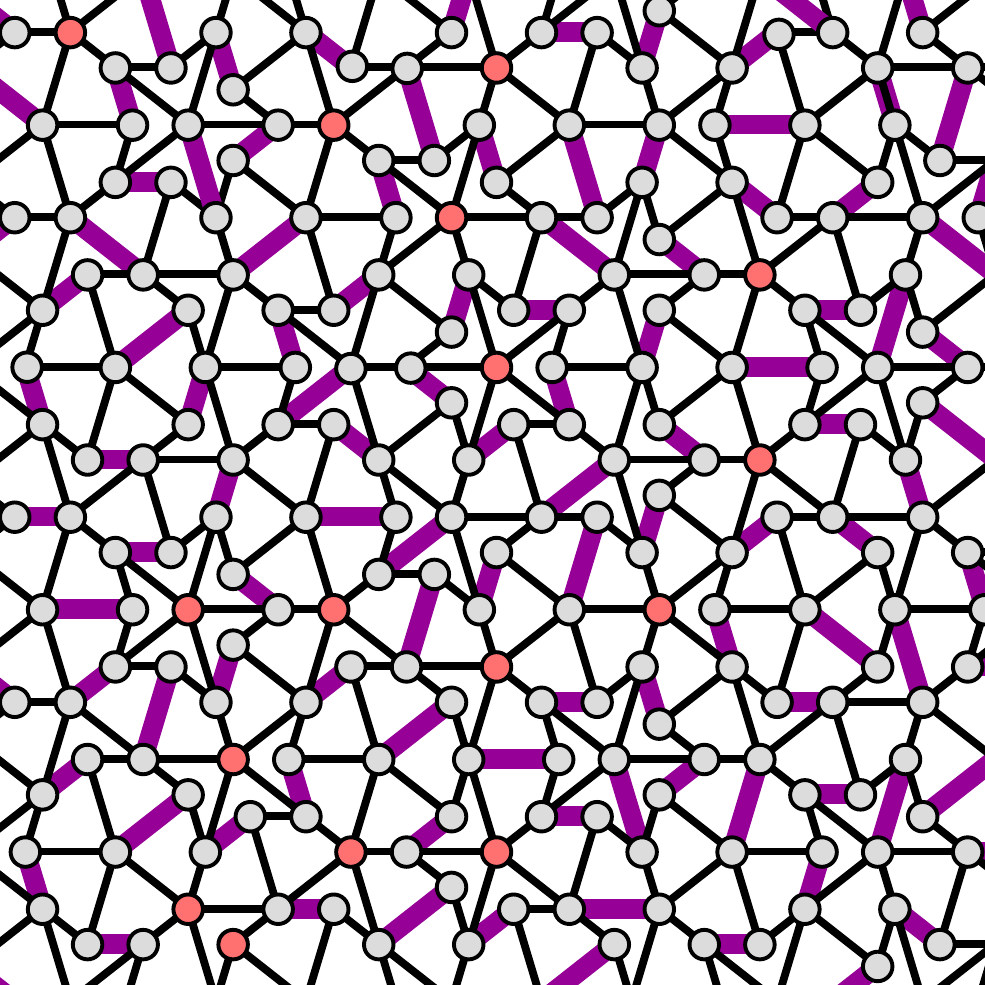}
\caption{A maximum matching of the Penrose dart-kite (P2) tiling in Fig.~\ref{fig:P2}. Monomers are shown in red. All lie on the same bipartite sub-lattice.
}
\label{fig:P2_matching}
\end{figure}

Fig.~\ref{fig:P2_matching} shows a maximum matching of the same region, obtained using the Hopcroft-Karp algorithm. A number of monomers appear, all on the same bipartite sub-lattice. Inspecting the vertex frequencies in Table~\ref{tab:frequencies} reveals an imbalance in the relative frequencies of vertices in the two sub-lattices: in the infinite tiling the Star-Queen-King-Ace sub-lattice contains more vertices than the Sun-Deuce-Jack sub-lattice. A perfect matching is therefore impossible, as the lower bound on the monomer density is given by the excess density of one sub-lattice over the other. In fact, this lower bound is the true monomer density. To see this, it suffices to prove that monomers on the Star-Queen-King-Ace sub-lattice can reach any vertex on that sub-lattice: if this is the case, the Sun-Deuce-Jack sub-lattice must be perfectly matched, otherwise the total number of monomers could be decreased by moving a monomer on the Star-Queen-King-Ace sub-lattice next to a monomer on the Sun-Deuce-Jack sub-lattice, then annihilating both. Therefore, if the monomers present on the Star-Queen-King-Ace sub-lattice can reach any vertex of that sub-lattice, then the Sun-Deuce-Jack sub-lattice is perfectly matched, and the matching contains the minimum number of monomers.

{\bf Any vertex on the Star-Queen-King-Ace sub-lattice can be reached by a monomer}. \emph{Proof}: by considering an arbitrary maximum matching of the cartwheel with each of its continuations, any vertex on the Star-Queen-King-Ace sub-lattice can be seen to connect via an alternating path to a monomer. As every vertex in the infinite tiling appears within a cartwheel, every vertex on the infinite Star-Queen-King-Ace sub-lattice connects to a monomer by an alternating path, and all vertices on this sub-lattice can be reached by a monomer. $\square$

Therefore the monomer density on P2 is precisely given by the imbalance of vertices between the two sub-lattices. This can be found as follows:
\begin{align}
\rho_{\textrm{monomer}}=\frac{\sum_{v\in A}f\left(v\right)-\sum_{v\in B}f\left(v\right)}{\sum_{v\in A}f\left(v\right)+\sum_{v\in B}f\left(v\right)}
\end{align}
where $f\left(v\right)$ is the relative frequency with which vertex $v$ appears (see Table~\ref{tab:frequencies}), $A$ denotes the set of vertices on the Star-Queen-King-Ace bipartite sub-lattice, and $B$ denotes the set of vertices on the Deuce/Jack/Sun sub-lattice. The result simplifies to 
\begin{align}
\rho_{\textrm{monomer}}=\left(7-4\varphi\right)/5.
\end{align}
We again confirmed this result numerically by employing the Hopcroft-Karp algorithm to find maximum matchings of successively larger finite-sized regions created by inflating the two basic tiles. The result, shown in Fig.~\ref{fig:numerical_density}, shows a rapid convergence towards the analytic result, which is valid in the thermodynamic limit.

\subsubsection{Decorated Penrose Tilings}

\begin{figure}[t]
\includegraphics[width=.4\textwidth]{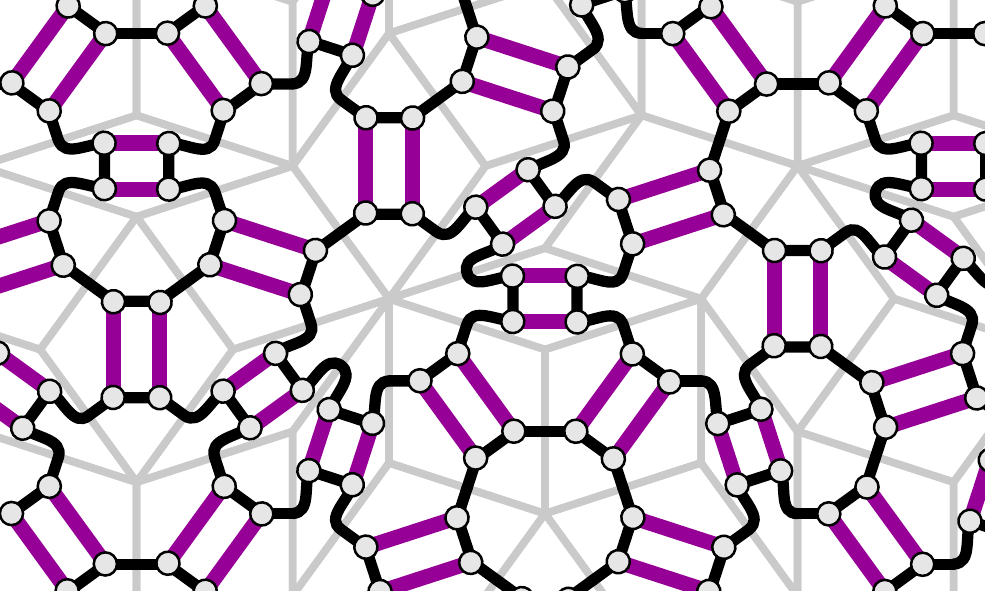}
\caption{The rhombic Penrose tiles (light grey) decorated with edges (black), vertices, and dimers (purple) such that each tile of the same type is identical and the resulting bipartite graph admits a perfect matching.
}
\label{fig:perfect}
\end{figure}

The different Penrose tilings P1--P3 can be derived from one another as decorations of the tiles. P1 is non-bipartite, P2 is bipartite and features charged maximum matchings, and P3 is bipartite and features charge-neutral maximum matchings. It is natural to ask whether a decoration of the tiles is possible which leads to a bipartite graph admitting perfect matchings. In fact such decorations are simple to design, with one option shown in Fig.~\ref{fig:perfect}. Each tile of P3 has been decorated with edges, vertices, and dimers, in such a way that a perfect matching is present by construction~\footnote{This design was developed during discussions with N.~G.~Jones}. All vertices of P3 appear in the region shown, and it can be seen that the number of graph edges enclosing a valence-$v$ vertex is $2v$. As this number is even, and the only other cycles (appearing within the tiles) are of length four, the graph is bipartite. Some of the edges of the graph have been allowed to curve for ease of drawing. 

\subsection{Eightfold and Twelvefold Penrose-like Tilings}
\label{subsec:8_and_12}

By considering various decorations of Penrose tilings we have found a wide range of behaviors of classical dimer models. In order to establish the generality of these results, in this section we briefly consider classical dimers on other two-dimensional Penrose-like tilings. 

All known examples of physical quasicrystals have 5-, 8-, 10-, or 12-fold rotational symmetry, and feature symmetries related to quadratic-irrational PV numbers~\cite{Levitov88,BoyleSteinhardt16A,ZaporskiFlicker18}. In the present paper we have constructed Penrose-like tilings using inflation rules. An alternative construction method is based on the use of \emph{Ammann bars}, decorations of the tiles with straight line segments such that valid tile configurations lead to infinite sets of straight lines~\cite{Gardner,GrunbaumShephard}. Reference~\onlinecite{BoyleSteinhardt16B} identifies that there are only six possible two-dimensional Penrose-like tilings which have the minimal set of one-dimensional Ammann decorations compatible with their orientational symmetries. The authors of that paper provide a method of constructing the tilings from the Ammann bars. The construction leads to tilings made up of small numbers of inequivalent convex tiles (as in P3, which has two tiles). The six two-dimensional Penrose-like tilings resulting from this construction are as follows: the rhombic Penrose tiling (with a 10-fold symmetric diffraction pattern, and PV number $\varphi$), tilings 8A and 8B with 8-fold symmetric diffraction patterns and PV number $1+\sqrt{2}$ (the silver ratio), and tilings 12A, 12B, and 12C with 12-fold symmetric diffraction patterns and PV number $2+\sqrt{3}$. 

We save a full analysis of the behavior of dimer matchings in all of these cases for future work. However, we are able to make a number of comments. Case 8A is better known as the Ammann-Beenker tiling~\cite{GrunbaumShephard,BaakeGrimm}. In upcoming work we prove that this case admits perfect matchings~\cite{LloydEA19}. Turning to case 8B, we were able to find a perfect matching of a large finite region, and we conjecture that the infinite tiling also admits perfect matchings. 

\begin{figure}[t]
\includegraphics[width=.4\textwidth]{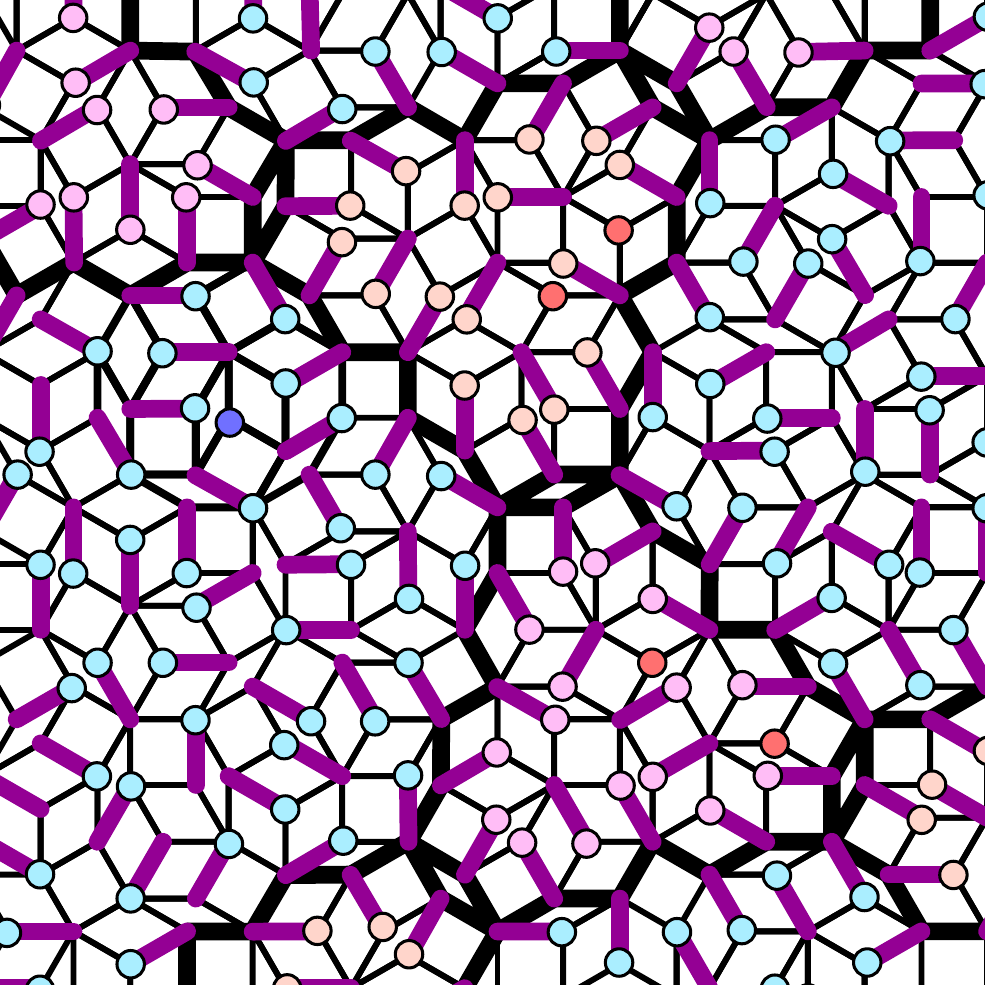}
\caption{A maximum matching of the Socolar-Taylor tiling~\cite{Socolar89,BoyleSteinhardt16B}. Purple edges indicate dimers; the dark-blue (four dark-red) vertices indicate monomers, while light-blue (light-red) vertices connect to blue (red) monomers via alternating paths (note the two distinct colors of light red, indicating distinct monomer regions; the separation of the different blue regions is clear). Monomer membranes have been identified with thick black lines. 
}
\label{fig:12A}
\end{figure}

Case 12A is better known as the Socolar-Taylor tiling~\cite{Socolar89,BoyleSteinhardt16B}. It contains three tiles with angles which are multiples of $2\pi/12$. Creating a maximum matching of a finite region, we found that monomers of both charges necessarily exist, and are trapped by uncrossable membranes as in the rhombic Penrose tiling. A sample of this matching is shown in Fig.~\ref{fig:12A}. The existence of any monomer unable to reach the boundary of a finite region, in any maximum matching, is sufficient to prove that perfect matchings do not exist for the infinite tiling. The matching in Fig.~\ref{fig:12A} contains one blue monomer and four red monomers; in lighter colors are vertices which connect to these monomers via alternating paths, and which can therefore be reached under minimal monomer moves. While the matching itself is arbitrary, the set of vertices which can be reached by monomers is independent of matching~\cite{LovaszPlummer}. Monomer membranes separate regions of opposite bipartite charge, as in the rhombic Penrose tiling, and again appear to follow chains of 4- and 6-vertices interspersed with 5-vertices. Unlike the Penrose tiling, however, the Socolar-Taylor membranes are able to branch. Branchings appear to occur at double-width membrane-segments formed from two 4-vertices appearing back-to-back across a thin tile. These double-width segments are then able to separate regions of the same bipartite charge (we have used two different colors of light red to indicate these distinct regions). 

\begin{figure}[t]
\includegraphics[width=.4\textwidth]{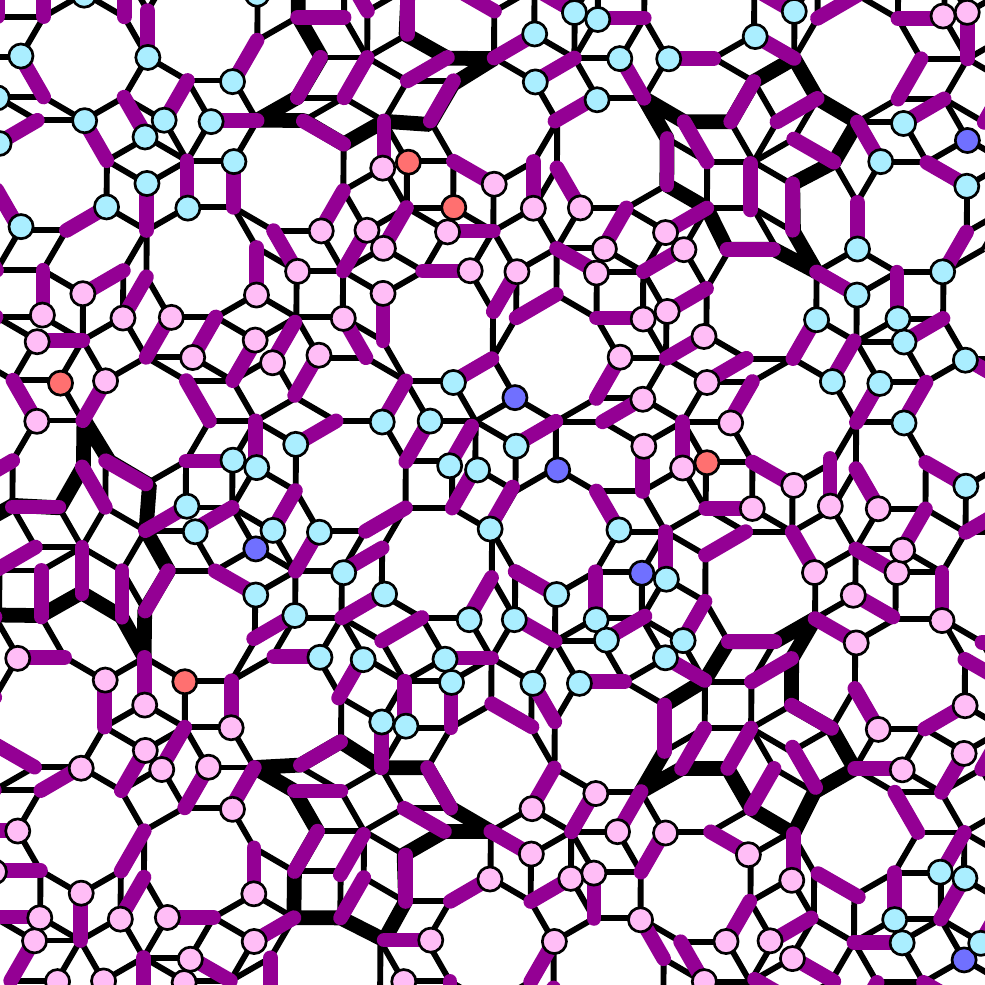}
\caption{A maximum matching of the Boyle-Steinhardt 12B tiling~\cite{BoyleSteinhardt16B}. Colors as in Fig.~\ref{fig:12A}. Note the existence of perfectly-matched regions within monomer membranes, thick black lines. These resemble manta rays converging on the center. These perfectly-matched regions appear to be unique to the 12B and 12C tilings. 
}
\label{fig:12B}
\end{figure}

Cases 12B and 12C are similarly unable to admit perfect matchings, and feature both single- and double-width membranes as in 12A. Whereas all monomer regions we identified in 12A contained a net imbalance of charge, as in the Penrose tiling, cases 12B and 12C additionally feature perfectly-matched islands enclosed by membranes. A maximum matching of a region of 12B is shown in Fig.~\ref{fig:12B}.

%
\section{Concluding Remarks}
\label{sec:conclusions}
%

In this paper we have considered dimer coverings of the Penrose tiling, considered as a graph. We found that a perfect matching is not possible, but identified various properties of maximum matchings, which are those that have the largest possible number of dimers (smallest number of monomers). We devised a method of generating these maximum matchings using the properties of the Penrose tiling (the \emph{dimer inflation algorithm}), and used this to show that the density of monomers in such matchings is $81-50\varphi$. These monomers exist in closed \emph{monomer regions}, bounded by loops of second-nearest neighbor even-valence vertices. These loops are fractal objects, that we dubbed \emph{monomer membranes}, which the monomers cannot cross. Each monomer region has an excess of one or other bipartite charge, equal to the number of monomers it contains in maximum matchings. Regions on opposite sides of a membrane have opposite net bipartite charge. 

We note that an immediate  physical application of our results is as a model for adsorbed atoms and molecules on the surfaces of quasicrystals, already known to lead to a variety of exotic structures~\cite{McGrathEA02,Trasca04,McGrath12}; it is straightforward to translate our analysis into a series of statements about such structures. However, as noted in the introduction, the dimer model can be used to study a wide range of phenomena, and therefore we might anticipate various other possible applications, as we now sketch. 

First, we note that the physics of local constraints can drive a variety of rich phenomena even at the classical level. Perhaps the most famous recent example is the study of condensed-matter analogues of \emph{magnetic monopoles}~\cite{CastelnovoEA09,MorrisEA09,BramwellEA09}. Although they have eluded detection as fundamental particles, various experiments have indicated that they may emerge as collective excitations in the spin-ice materials dysprosium titanate and holmium titanate (A$_2$Ti$_2$O$_7$ with A = Dy, Ho)~\cite{CastelnovoEA09}. In these materials, the magnetic rare-earth ions inhabit a pyrochlore lattice of corner-sharing tetrahedra~\footnote{Alternatively, we can view them as forming a face-centred cubic lattice with a four-site basis.}. By an appropriate choice of local spin axes, the low-energy physics may be approximately captured by that of a nearest-neighbor Ising antiferromagnet. In any classical ground state, two of the four spins on a tetrahedron point into the centre, and two point out. Since there are six possible configurations per tetrahedron, the result is a macroscopically degenerate ground state characterized by the local two-in, two-out \emph{ice-rule} constraint~\cite{Gingras09}. This can be viewed as a Gauss' law for an emergent gauge field, but is also linked to the local magnetization since each spin is a magnetic dipole. The lowest-energy excitation out of the ice-rule manifold consists of a single spin flip, which creates a three-in one-out tetrahedron neighboring a one-in three-out. Upon coarse-graining by summing the divergence of the magnetization over each tetrahedron, the excitation appears as a neighboring source and sink of magnetization, termed a monopole and anti-monopole (the analogue to monomers and anti-monomers in the case of the dimer model). Subsequent spin flips allow the monopole and anti-monopole to move apart. A Dirac string of flipped spins connects monopole anti-monopole pairs, analogous to augmenting paths in the Penrose tilings considered here. In the spin-ice setting, statistical fluctuations between different spin configurations make the precise location of the string ambiguous (except with reference to a chosen starting state), giving a classical analogue of the underlying gauge theory~\cite{MorrisEA09}. The strings also lead to an effective Coulomb's law interaction between monopoles. This picture is a clear demonstration of the idea of fractionalization, albeit in a classical context, and its parallels to dimer models are clear. Our results suggest that classical frustrated magnets in quasiperiodic systems should host a similarly rich set of phenomena.

Quantum fluctuations between classically degenerate spin ice configurations --- generated, for instance, by transverse-field terms beyond the Ising limit --- can give rise to a quantum gauge theory in its Coulomb phase, with a gapless `photon' collective mode. In such a phase, often termed a three-dimensional $U(1)$ quantum spin liquid, the monopoles are emergent gapped quasiparticles, which carry a $U(1)$ charge and exhibit a Coulombic interaction. The search for such {\it quantum spin ices} remains an active field of research~\cite{McClartyGingras14,SavaryBalents17}.

Similarly, endowing dimer configurations with dynamics --- the simplest of which is a \emph{resonance} that augments the elementary alternating cycle on a single four-site plaquette --- yields a quantum dimer model. On periodic bipartite lattices in three dimensions, such dimer models are also known to exist in a Coulomb liquid phase~\cite{MoessnerSondhiRVB_3D,MoessnerRaman07}; here, the `photon' is a collective mode of the dimers, whereas the monomers are gapped, gauge-charged objects (similar to the monopoles in spin ice). However, on periodic non-bipartite lattices in any dimension, quantum dimer models lead to fully gapped dimer liquid phases with a discrete gauge structure and topological order~\cite{MoessnerSondhi01,MoessnerRaman07}. The situation is more subtle on two-dimensional periodic bipartite lattices, as in the case of the original square-lattice quantum dimer model~\cite{RokhsarKivelson88}. On such lattices, although classical dimers exhibit power-law correlations, the corresponding quantum Coulomb liquid phase is generically destroyed by instanton effects~\cite{Polyakov1977}. These lead to the formation to a variety of dimer crystal phases, in which the emergent gauge field is confined. Deconfinement (as in the Coulomb phase) only survives at critical points between these crystal phases. However, more careful analysis of the effective theory near such transitions reveals that lattice effects can have a significant impact on their properties~\cite{FradkinEA04,VBS04}. If the incipient crystalline order is incommensurate with the underlying periodic lattice, the dimer model remains gapless and the collective mode survives. The interplay between the different possible dimer crystal orders and the lattice leads to a \emph{Devil's staircase} of gapped commensurate crystals interleaved with gapless incommensurate ones --- a phenomenon dubbed \emph{Cantor deconfinement}~\cite{FradkinEA04}. This provides one obvious motivation to study quantum dimer models on lattices, such as the Penrose tiling, where any crystalline order is likely to be frustrated.

The presence of a finite local density of monomers (but vanishing net monomer charge) suggests a route to evading the effects of instantons in a quantum extension of the present model: the presence of dynamical gauge-charged matter is known to mitigate their influence~\cite{HermeleEA04,RyuEA07}. Intriguingly, the presence of monomer membranes that are impermeable to gauge charge suggests a rather unusual phase structure, that blends aspects both confinement and deconfinement. The presence of fractal structures that constrain the low-energy dynamics also bears a family resemblance to the physics of \emph{fractons}. These are immobile quasiparticles that appear as low-energy excitations in certain translation-invariant Hamiltonian models~\cite{Chamon2005,bravyi2009no,Haah2011,NandkishoreHermele18}, which may be viewed as end points of fractal objects. The immobility of fractons and their glassy dynamics~\cite{Chamon2005,GlassyFractonDynamics} is closely related to the properties of simple classical spin models with kinetic constraints~\cite{NewmanMoore}. While the precise connection between our work, the physics of fracton models, and these related physical systems remains unclear, we flag this as an interesting avenue for further study.

The various lines of investigation suggested above would clearly be advanced by the development of a convenient coarse-grained description of the Penrose dimer model. For dimer models on 2D periodic lattices, such a description is provided by a mapping to a so-called {\it height model} that parameterizes dimer configurations in terms of configurations of a two-dimensional surface or height field. 
A conventional scalar height model cannot readily be defined on the Penrose tiling owing to the variation in valence of the various vertices. Suitable generalizations may exist which achieve the same result.

In a different vein, it would be interesting to determine which of the properties we have demonstrated in the Penrose tiling carry over to other Penrose-like tilings, the members of the non-crystallographic Coxeter groups~\cite{BoyleSteinhardt16B,Coxeter35}. There are an infinite number in 1D (\emph{e.g.} the Fibonacci quasilattice~\cite{GrunbaumShephard,Flicker18,ZaporskiFlicker18}), six in 2D (considered in Section~\ref{sec:other_quasilattices}), five in 3D (with the point group symmetries of the icosahedron and dodecahedron), one in 4D (with the point group symmetry of the 600-cell), and none in dimension five or higher. All can be generated by inflation rules, matching rules, and by a cut-and-project method from higher dimensions, suggesting that similar methods to those we have developed here may be extended to those cases. The three-dimensional cases in particular would be interesting candidates for physically-relevant systems exhibiting topological order. We also hope that the present physically-motivated study of the Penrose tiling may open up new directions for studying the fascinating mathematical properties of these tilings, such as their three-colorability~\cite{Gardner,GrunbaumShephard,SibleyWagon00}. 

Part of the interest in dimer coverings of graphs lies in their potential relevance to the problem of discretizing conformal field theories governing critical systems~\cite{Kenyon99,Kenyon02,ArdonneEA04}. As Penrose-like tilings lack the discrete translational invariance of periodic lattices, but instead feature a discrete scale invariance, they seem a natural subject of study from this perspective. Indeed,   
in the special case of \emph{conformal quasicrystals}, which form the boundary of regular tilings of hyperbolic space, the structures have invariance under discretized Weyl transformations, and so can be considered to host a full conformal invariance~\cite{BoyleEA19}.

We anticipate that the themes explored here will prove relevant to understanding the range of emergent strongly correlated phenomena possible in quasicrystals. Our aim in presenting these results has been to lay the foundations for future investigations into new and unconventional forms of classical and quantum order possible in these systems.

%
\section*{Acknowledgments}
%
We thank Sounak Biswas, Claudio Castelnovo, Kedar Damle, Nick Jones, Jerome Lloyd, Roderich Moessner, and Jasper van Wezel for useful discussions. F.F.~acknowledges support from the Astor Junior Research Fellowship of New College, Oxford. S.H.S.~was supported by EPSRC grant EP/N01930X/1. S.A.P.~acknowledges support from the European Research Council (ERC) under the European Union Horizon 2020 Research and Innovation Programme (Grant Agreement No.~804213-TMCS). Statement of compliance with EPSRC policy framework on research data: This publication is theoretical work that does not require supporting research data. 

\bibliographystyle{apsrev4-1}

%

%
\newpage
\pagebreak
\appendix

\counterwithin{figure}{section}

%
\onecolumngrid
\section{Proof monomers cannot reside on monomer membranes}
\label{app:even_loop_proof}
%

In this appendix we prove that monomers cannot appear on 4-vertices, or on $5_C$-vertices appearing between 4-vertices and 6-vertices, in a maximum matching. Collectively these vertices comprise closed loops which monomers cannot cross, and which we refer to as monomer membranes. Note that monomers \emph{can} appear on $5_C$-vertices appearing between two 4-vertices, but only in the exceptional case of the $4^5$ loop (see Section~\ref{sec:no_perfect_matchings}). 

\begin{figure*}[t]
\includegraphics[width=\textwidth]{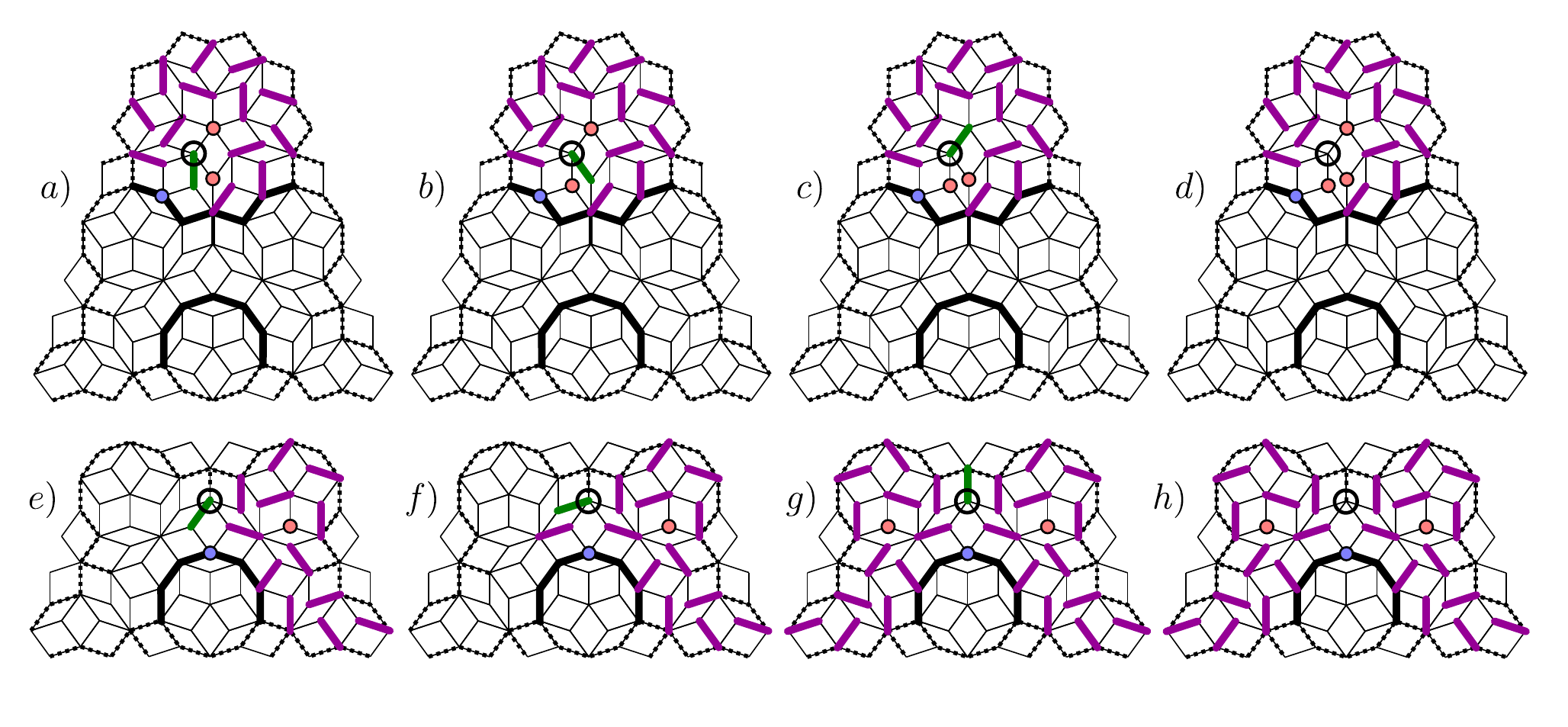}
\caption{Proof no monomer can be based on a 4-vertex in a maximum matching. See accompanying text in Appendix~\ref{app:even_loop_proof}.
}
\label{fig:no_monomer_4}
\end{figure*}

In Fig.~\ref{fig:no_monomer_4} we prove that no monomer can appear on a 4-vertex in a maximum matching. The local empire of the 4-vertex is too small to construct a direct analogue to the argument presented for the 6-vertex. However, 4-vertices have either 4-vertices or 6-vertices as second-nearest neighbors (with $5_C$ appearing as the connecting vertex). Figs.~\ref{fig:no_monomer_4}(a)-(d) assume a $-\boxed{4}-6-$ configuration, where the boxed symbol is the vertex under consideration, and the $5_C$ connecting vertices are not listed. Figs.~\ref{fig:no_monomer_4}(e)-(h) assume a $-4-\boxed{4}-4-$ configuration. Place a monomer (blue) on the 4-vertex. The circled second-nearest neighbor can host dimers in one of three symmetry-inequivalent positions (a)-(c), or it can host a monomer (d). In all cases, it can be seen that at least two further monomers are implied which connect via augmenting paths to the monomer on the 4-vertex (a contradiction). In (e)-(h) the same argument applies in the other configuration. Therefore a monomer cannot appear on a 4-vertex in a maximum dimer matching. $\square$

\begin{figure}[t]
\includegraphics[width=.5\textwidth]{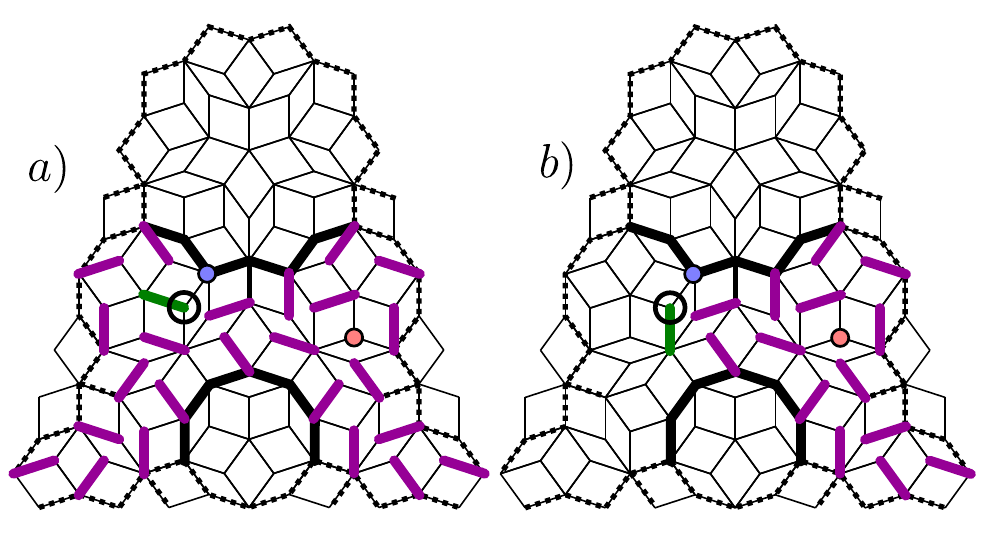}
\caption{Proof no monomer can be based on a $5_C$-vertex appearing between a 4-vertex and a 6-vertex in a maximum matching. See accompanying text in Appendix~\ref{app:even_loop_proof}.
}
\label{fig:no_monomer_5c}
\end{figure}

In Fig.~\ref{fig:no_monomer_5c} we prove that no monomer can be based on a $5_C$-vertex appearing between a 4-vertex and a 6-vertex in a maximum matching. Place a monomer (blue) on the $5_C$-vertex. The circled vertex must host a dimer if an augmenting path is to be avoided. The two options are shown in (a) and (b), and in both cases the second monomer is shown to be unavoidable. $\square$

\begin{figure*}[t]
\includegraphics[width=\textwidth]{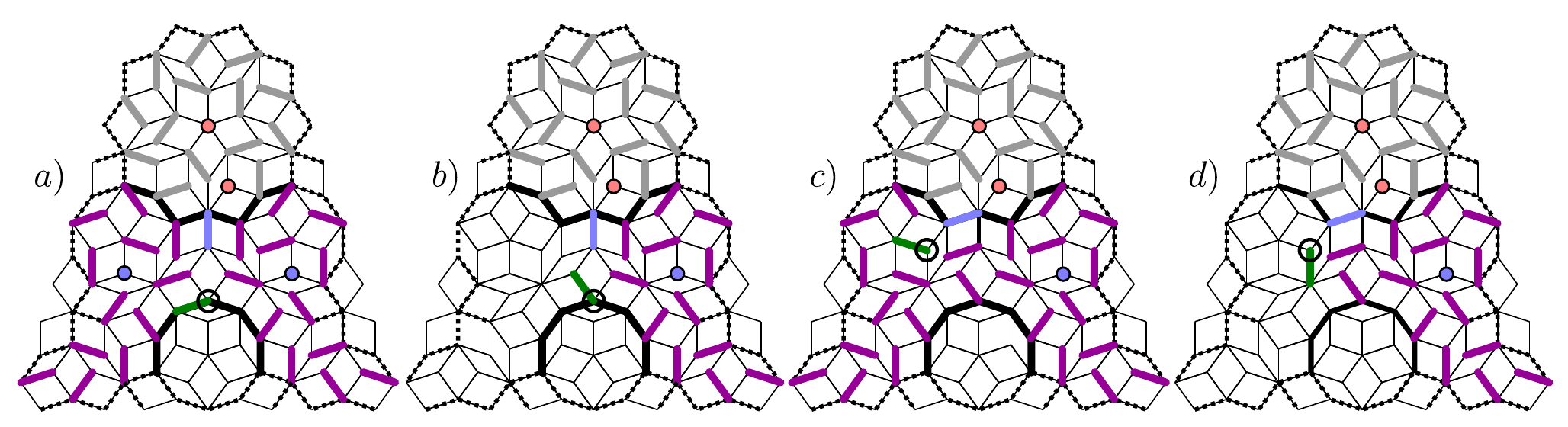}
\caption{Proof that, of the edges emanating from a 6-vertex, only the edges indicated in Fig.~\ref{fig:even_loop_proof}b may host dimers. See accompanying text in Appendix~\ref{app:even_loop_proof}.
}
\label{fig:no_legs_6}
\end{figure*}

In Fig.~\ref{fig:no_legs_6} we prove that the dimer which must connect to the 6-vertex must appear on one of the three legs indicated in Fig.~\ref{fig:even_loop_proof}b. The proof takes the following form. Based on the results of Figs.~\ref{fig:even_loop_proof} and \ref{fig:no_monomer_4}, no monomer can exist on 4-vertices or 6-vertices. The proofs in those figures also demonstrate that if the monomer is substituted with a dimer extending into the region on the opposite side of the thick black line, augmenting paths can exist which include this dimer. After augmentation the dimer will be returned to one of the edges specified in Fig.~\ref{fig:even_loop_proof}b. Rephrasing in terms of minimal monomer moves, exactly one monomer will be able to cross the black line. This cannot occur in a maximum matching if there is a monomer of opposite bipartite charge waiting on the other side of the thick black line, as the two monomers will then be able to annihilate, and should not have been present in a maximum matching. In Figs.~\ref{fig:no_legs_6}(a) and (b), above the thick black line we reproduce one of the maximum matchings from Fig.~\ref{fig:even_loop_proof}a, with the dimers recolored to grey to indicate that the specific configuration is unimportant to the argument (any case will work). We assume the 6-vertex does not host a monomer, but instead hosts the forbidden dimer indicated in blue. By considering the possible edges of the circled 4-vertex directly below, of which there are two symmetry-inequivalent choices, we show the implied dimers in each case, and in each case a blue monomer is implied with the opposite charge to the monomers in the region above the line. As the blue dimer connects the regions, there exists an augmenting path (\emph{i.e.}~exactly one monomer can cross the wall to annihilate), and the matching was not maximum. Note that the choice of the circled 4-vertex was simply out of convenience, and other vertices could have been considered. In (b) we construct a similar proof for the remaining symmetry-inequivalent edge of the 6-vertex, again covered by a dimer indicated in blue, and this time the circled 3-vertex is convenient to consider. $\square$

\begin{figure*}[t]
\includegraphics[width=\textwidth]{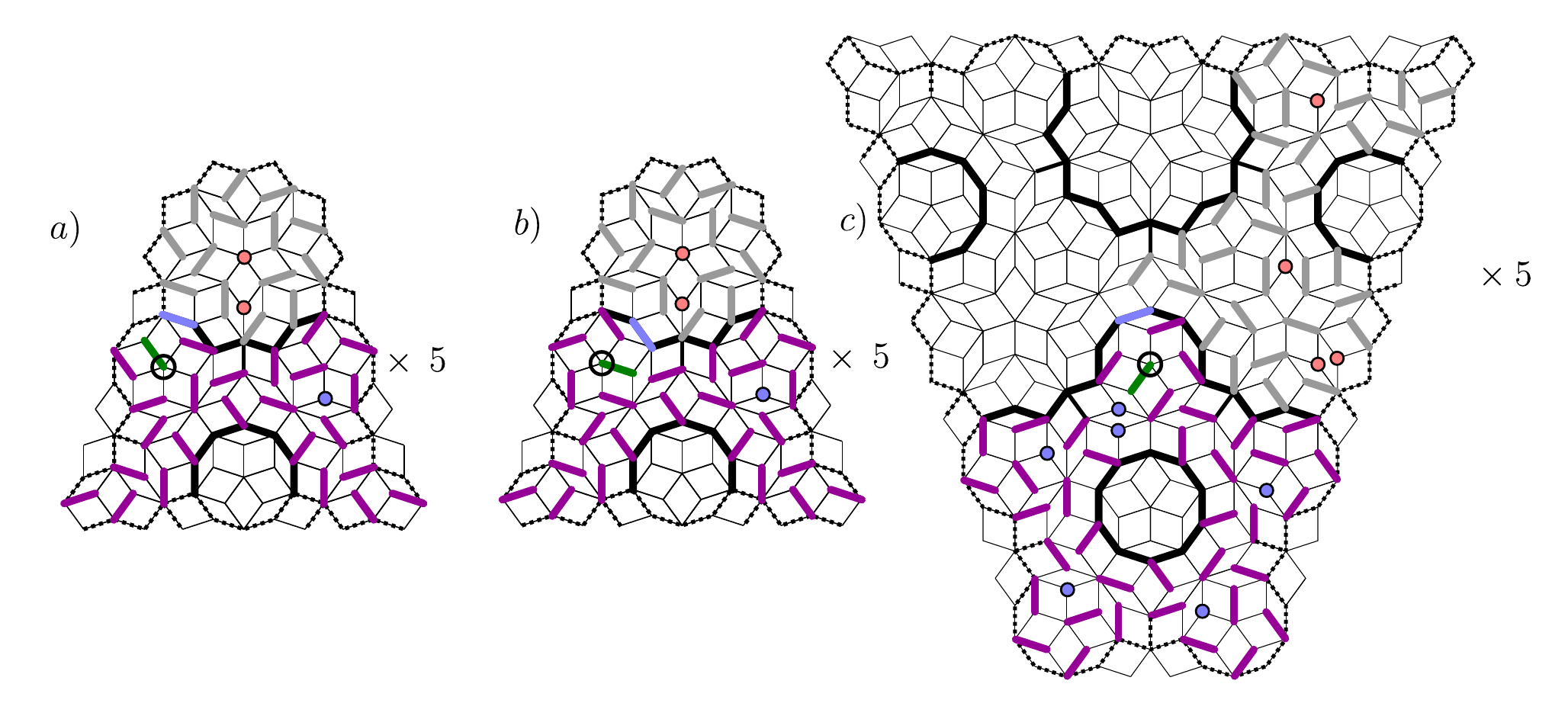}
\caption{Proof that, of the edges emanating from a 4-vertex, only the edges indicated in Fig.~\ref{fig:even_loop_proof}b may host dimers. See accompanying text in Appendix~\ref{app:even_loop_proof}, and Fig.~\ref{fig:no_legs_6}. 
}
\label{fig:no_legs_4}
\end{figure*}

In Fig.~\ref{fig:no_legs_4} we prove that the dimer which must connect to the 4-vertex must appear on one of the two legs indicated in Fig.~\ref{fig:even_loop_proof}b. The case of the 4-vertex is complicated by the relatively smaller local empire of the 4-vertex. Three cases need to be considered, recalling that any even-valence vertex has two even-valence vertices as second-nearest neighbors. First, the 4-vertex may have a 6-vertex as a second-nearest neighbor, as in Figs.~\ref{fig:no_legs_4}(a) and (b). In this case we may again consider the local empire of the 6-vertex. We place the blue dimer on each of the edges in question, and we consider a convenient nearby vertex. In this case the nearby $5_A$-vertex works well (shown). The five edges connected to this vertex must be considered, of which only one choice is shown in (a) and (b). The other cases can be checked quickly and yield the familiar result, that the forbidden placement of the blue dimer allows a single monomer to cross the black line and annihilate with an oppositely-charged monomer, so this configuration will not appear in a maximum matching. The second option is that the 4-vertex has 4-vertices on either side, but that the vertex after that is a 6-vertex. The local empire of this configuration is shown in (c); in fact, it implies the chain $-4-6-4-\boxed{4}-4-6-4-$ as shown. There is only one symmetry-inequivalent edge to consider; choose again the circled neighboring $5_A$-vertex, and any of the five dimer placements implies blue monomers to annihilate with the red monomers in the other region. The final option is that the 4-vertex appears in a $4^5$ ring (also present elsewhere in (c)). In this case the proof cannot be constructed, as this is the only case in which the otherwise-forbidden edges of the 4-vertex \emph{can} be covered by a dimer. Only one of the five such edges may be covered by a dimer, and this is only if there is no monomer within the ring. 

Finally, note that when a $5_C$-vertex appears as part of an impermeable monomer membrane (\emph{i.e.}~any membrane larger than $4^5$), the set of edges which can be covered by dimers is restricted. The easiest way to state the restriction is that the edges which cannot be covered are those which are either already implied by the uncoverable edges of 4- or -6-vertices, and the edge lying along the mirror plane of the $5_C$-vertex whenever deletion of this edge will lead to a disconnected graph. In the $-4-\boxed{5_C}-6-$ configuration this result has been proven as part of the proofs just given for the 4- and 6-vertices (the relevant uncoverable legs of the $5_C$-vertex \emph{are} the uncoverable legs of the 4- or 6-vertex). In the $4^5$ configuration the result does not hold, and all legs of the $5_C$-vertex are potentially coverable. The only remaining case is the $-4-6-4-\boxed{5_C}-4-4-6-4-$ configuration (where only the relevant $5_C$ vertex is listed). Two of the uncoverable legs are again proven uncoverable by the 4-vertex proof. The remaining case also follows directly from Fig.~\ref{fig:no_legs_4}c.

%
\section{Proof that the dimer inflation algorithm generates maximum matchings}
\label{app:dimer_inflation_proof}
%

In this appendix we prove the following statements made in Section~\ref{subsec:dimer_inflation_proof}:
\begin{itemize}
\item any path connecting any two 7-vertices is of even (odd) length if it crosses impermeable monomer membranes an even (odd) number of times
\item any path connecting any two $5_{A,B}$-vertices, where the $5_{A,B}$-vertices have no 7-vertices as second-nearest neighbors, is of even (odd) length if it crosses impermeable monomer membranes an even (odd) number of times
\item any path connecting any 7-vertex to any $5_{A,B}$-vertices, where the $5_{A,B}$-vertices have no 7-vertices as second-nearest neighbors, is of odd (even) length if it crosses impermeable monomer membranes an even (odd) number of times.
\end{itemize}
Fig.~\ref{fig:algorithm_maximum_proof} shows the local empire of the 6-vertex, sufficient to cover the Penrose tiling. Coloured disks indicate vertices of certain valences which are either definite (solid circles) or potential, depending on the surrounding patches (dashed circles). Red vertices are 7-vertices; blue vertices are $5_A$-vertices; pink vertices are $5_B$-vertices. The thick solid line, indicating the monomer membrane boundary passing through the 6-vertex, disconnects vertices $a-g$ plus the internal red and blue vertices, from vertices $h-k$ and the internal pink vertex, in all possible continuations of the monomer membrane. Note that all internal vertices obey the specified rules, and so, when considering a boundary vertex, its relationship only needs to be shown to be correct to any one internal vertex.
\begin{itemize}
\item red vertices $c$, $e$, and $g$ are connected by even-length paths to the interior red vertices $\checkmark$
\item all these vertices are connected by odd-length paths to red vertices $h$ and $j$, which must be separated by a monomer membrane (if $h$ is a 7-vertex it implies $g$ is not, as the potential monomer membrane is resolved to pass downwards, implying $g$ is a $5_C$-vertex) $\checkmark$
\item red vertex $b$ is connected by odd-length paths to the interior red vertices. However, if $b$ is a 7-vertex, this forces $c$ (and its mirror-equivalent) to be a 6-vertex, in which case $b$ is separated from the interior red vertices by an impermeable monomer membrane $\checkmark$
\item blue vertex $a$ is connected to the interior blue vertices by odd-length paths. If $b$ is a 4-vertex then $a$ is a $5_A$-vertex with no 7-vertices as second-nearest neighbors, so $a$ receives a monomer of the same bipartite charge, and since the interior blue vertices have at least one 7-vertex as a second-nearest neighbor they receive monomers of opposite bipartite charge to themselves, \emph{i.e.}~the same as the charge of $a$. This is correct. If $b$ is not a 4-vertex it must be a 7-vertex. The monomer associated to $a$ is now of the opposite bipartite charge to $a$; but if $b$ is not a 4-vertex then $c$ must be a 6-vertex to continue the monomer membrane, and so $a$ is separated by an impermeable monomer membrane from the interior blue vertices. This too is correct $\checkmark$
\item blue vertex $d$ is connected by even-length paths to the internal blue vertices $\checkmark$
\item pink vertex $f$ is connected by odd-length paths to the internal red vertices $\checkmark$
\item red vertex $h$ is connected by odd-length paths to the internal pink vertex $\checkmark$
\item blue vertex $i$ is connected by odd-length paths to the internal pink vertex. If $i$ is a $5_A$-vertex then $h$ and $j$ must be 7-vertices, and so the monomers associated to $i$ are of opposite bipartite charge to $i$. As the monomer associated to the pink vertex is of the same charge as the pink, this is correct. If $i$ is a 4-vertex $h$ must be a 6-vertex, and there is no issue with the presence of the internal pink vertex $\checkmark$
\item similarly, if either $h$ or $j$ is a 7-vertex this forces the $i$ to be a $5_A$-vertex, and the same cases hold. The same cases also hold for $k$ $\checkmark$
\end{itemize}
This concludes the check of every possible boundary vertex of interest on the local empire of the 6-vertex (the unmarked boundary vertices either cannot be $5_A$-, $5_B$-, or 7-vertices, or the results are implied by the vertical mirror plane passing through the 6-vertex).


\end{document}